\documentclass[orivec, runningheads]{llncs}
\pdfoutput=1
\usepackage{cite}
\usepackage{amsmath}
\usepackage[cmintegrals]{newtxmath}
\usepackage[hyphens]{url}
\usepackage{graphicx}
\usepackage{algorithm}
\usepackage{algpseudocode}
\usepackage{subfigure}
\usepackage{diagbox}
\usepackage{multirow}
\begin{document}
%
% paper title
% Titles are generally capitalized except for words such as a, an, and, as,
% at, but, by, for, in, nor, of, on, or, the, to and up, which are usually
% not capitalized unless they are the first or last word of the title.
% Linebreaks \\ can be used within to get better formatting as desired.
% Do not put math or special symbols in the title.
\title{A Flexible Privacy-preserving Framework\newline
	for Singular Value Decomposition under\newline
	Internet of Things Environment}
%
%
% author names and IEEE memberships
% note positions of commas and nonbreaking spaces ( ~ ) LaTeX will not break
% a structure at a ~ so this keeps an author's name from being broken across
% two lines.
% use \thanks{} to gain access to the first footnote area
% a separate \thanks must be used for each paragraph as LaTeX2e's \thanks
% was not built to handle multiple paragraphs
%

\titlerunning{Privacy-preserving Framework for SVD under IoT}

\author{Shuo Chen\inst{1} \and Rongxing Lu\inst{2} \and Jie Zhang\inst{1} }

\authorrunning{Shuo Chen et al.}

\institute{Nanyang Technological University, Singapore, Singapore\\
	\email{chen1087@e.ntu.edu.sg},
	\email{zhangj@ntu.edu.sg}\\
	\and
	Faculty of Computer Science, University of New Brunswick, Canada\\
	\email{rxlu@ieee.org}}

%\author{\IEEEauthorblockN{Shuo Chen}
%	\IEEEauthorblockA{School of Computer Science and Engineering\\
%		Nanyang Technological University\\
%		Singapore\\
%		Email: chen1087@e.ntu.edu.sg}
%	\and
%	\IEEEauthorblockN{Rongxing Lu}
%	\IEEEauthorblockA{Faculty of Computer Science\\
%		University of New Brunswick\\
%		Canada\\
%		Email: rxlu@ieee.org}
%	\and
%	\IEEEauthorblockN{Jie Zhang}
%	\IEEEauthorblockA{School of Computer Science and Engineering\\
%		Nanyang Technological University\\
%		Singapore\\
%		Email: zhangj@ntu.edu.sg}}

%{{Shuo Chen},
%        Author B,~\IEEEmembership{Fellow,~OSA,}
%        and Author C,~\IEEEmembership{Life~Fellow,~IEEE}% <-this % stops a space

% make the title area
\maketitle

% As a general rule, do not put math, special symbols or citations
% in the abstract or keywords.
\begin{abstract}
The singular value decomposition (SVD) is a widely used matrix factorization tool which underlies plenty of useful applications, e.g. recommendation system, abnormal detection and data compression. Under the environment of emerging Internet of Things (IoT), there would be an increasing demand for data analysis to better human's lives and create new economic growth points. Moreover, due to the large scope of IoT, most of the data analysis work should be done in the network edge, i.e. handled by fog computing. However, the devices which provide fog computing may not be trustable while the data privacy is often the significant concern of the IoT application users. Thus, when performing SVD for data analysis purpose, the privacy of user data should be preserved. Based on the above reasons, in this paper, we propose a privacy-preserving fog computing framework for SVD computation. The security and performance analysis shows the practicability of the proposed framework. Furthermore, since different applications may utilize the result of SVD operation in different ways, three applications with different objectives are introduced to show how the framework could flexibly achieve the purposes of different applications, which indicates the flexibility of the design.
\end{abstract}

% Note that keywords are not normally used for peerreview papers.
%\begin{IEEEkeywords}
%singular value decomposition, fog computing, privacy-preserving, security, Internet of Things.
%\end{IEEEkeywords}

% For peer review papers, you can put extra information on the cover
% page as needed:
% \ifCLASSOPTIONpeerreview
% \begin{center} \bfseries EDICS Category: 3-BBND \end{center}
% \fi
%
% For peerreview papers, this IEEEtran command inserts a page break and
% creates the second title. It will be ignored for other modes.
%\IEEEpeerreviewmaketitle

\section{Introduction}
\label{sec1}
% The very first letter is a 2 line initial drop letter followed
% by the rest of the first word in caps.
% 
% form to use if the first word consists of a single letter:
% \IEEEPARstart{A}{demo} file is ....
% 
% form to use if you need the single drop letter followed by
% normal text (unknown if ever used by the IEEE):
% \IEEEPARstart{A}{}demo file is ....
% 
% Some journals put the first two words in caps:
% \IEEEPARstart{T}{his demo} file is ....
% 
% Here we have the typical use of a "T" for an initial drop letter
% and "HIS" in caps to complete the first word.
With the prosperous development of communication and computation technologies, the Internet of Things (IoT), which allows the physical objects to be sensed, accessed and controlled remotely through the network infrastructure, is no longer a fantasy nowadays. The big advantage of IoT is that with the data analysis on the huge amount of information collected from the physical world, the server is capable of making more accurate and optimal decisions which would produce considerable benefits. It is estimated that the global IoT market will be 14.4 trillion dollars by 2022 \cite{Bradley14}, the potential economic impact of IoT would be 3.9 to 11.1 trillion dollars by 2025 \cite{McKinsey2015}, and the number of devices connected to the Internet would be about 50 billion by 2020 \cite{Cisco2014}. It could be expected that rather than being conducted on the cloud or inside the intranet of companies, the data analysis work would be performed everywhere and anytime in the future due to the ubiquitous IoT. As the amount of data analysis tasks increases in IoT, the singular value decomposition (SVD), which is widely used in different data analysis applications  \cite{ide2004eigenspace}\cite{lee2013anomaly}\cite{sarwar2000application}\cite{billsus1998learning}\cite{kalman1996singularly}, will be performed frequently. However, the traditional way of performing SVD, i.e. calculating the SVD in central server, may not be practical in future IoT due to the huge amount of IoT devices. If all the data is transmitted to a central server for computation, it would lead to considerable computation and communication resource consumption of the server, which would further severly impact the quality of service (QoS) of IoT applications.

To ease the burden of the IoT server and guarantee the QoS, a new technique called fog computing, which is proposed by Cisco \cite{bonomi2012fog}, is suitable to be applied. The main idea of fog computing is to provide storage, computing and networking services between environmental devices and the central server. The fog devices which are in close proximity to end devices normally possess considerable storage and computation resource. With the equipped resource, the fog devices could process the collected data locally so as to loose the workload of the server. In specific, there are three tiers in the fog computing architecture: environmental tier, edge tier and central tier. In the environmental tier, there are billions of heterogeneous IoT devices collecting and uploading information of the physical world, e.g. medical sensors in eHealth and mobile phone of each people. The data collected by IoT devices will be transmitted to the edge tier. The distribution of fog devices in the edge tier are hierarchical which is a characteristic inherited from the traditional network architecture. For example, the switchers of a local area network could function as the first layer fog devices and the gateways which manage those switchers could serve as the second layer fog devices. The fog devices in the edge tier could perform the application-specific operations on received data locally and send the results to the server in the central tier. %The architecture of fog computing is shown in Fig. \ref{fig1}.
%Currently, the cloud computing which is equipped with giant computation resource and big data analysis technique is considered as a vital part of the future IoT. However, it is anticipated that there would be about 50 billion devices connected to the Internet by 2020 \cite{Cisco2014} which will collect or generate vast amount of data for analysis every moment while the structure of cloud computing is centralized. That means all the data need to be transmitted to the data center of cloud computing, which would cause a huge burden to the backbone network. 
%\begin{figure}[htbp]
%	\centering
%	\includegraphics[width=0.5\textwidth]{fig1}
%	%\makebox[\textwidth]{\includegraphics[width=\textwidth]{fig1}}\\
%	\caption{Architecture of Fog Computing}\label{fig1}
%\end{figure}
Owing to the processing of fog devices, the volume of data sent to server could be reduced to a large extent. Since the fog devices are spread in a highly distributed environment, it is impractical for the government or an institution which owns the central server to provide and maintain all those fog devices. Therefore, it is reasonable to assume that the fog devices would be supplied by third parties. %What's more, unlike the centralized cloud computing requiring high-performance computers which could only be provided by giant companies, the intelligent devices possessed by ordinary persons (e.g. smart phone, vehicle et al.) could also function as fog devices. 

Under the context of fog computing, one could perform the SVD operations on the fog devices. However, another problem which would appear is the privacy issue. The third parties which control the fog devices may not be trustworthy while in many IoT applications, the data collected from the environment is considered as private by the IoT application users, e.g. the vital signs in eHealth, the location and speed of vehicles, and the power usage in a smart grid. Performing the SVD on plaintext with fog devices is infeasible if the privacy is a primary concern from the perspective of data owners. Therefore, how to take advantage of fog computing to locally process data in a privacy-preserving way is a challenging issue.

In this paper, we propose a flexible fog computing framework for performing SVD with privacy preserved. The homomorphic encryption technique called Paillier encryption \cite{paillier1999public} is applied to protect the data privacy. The framework is designed to be capable of supporting different applications based on the SVD computation. The main contributions of this paper are three-fold.

\begin{itemize}
	\item First, to perform data analysis for IoT applications, we propose a fog computing framework for SVD computation to ease the burden of server. Since the computation is performed in the fog devices which may not be trustable, the framework achieves the privacy-preserving SVD computation.
	\item Second, there is only one communication round between the data providers and data processors in our work while most of the existing works require iterative communications, which brings heavy overhead.
	\item Third, besides the framework for basic SVD operation, three applications are introduced in details to demonstrate the flexibility of the framework. It has been shown that the proposed framework could flexibly adapt to different applications with slight adjustments or several extra procedures.
\end{itemize}

The reminder of this paper is organized as follow. In Section \ref{sec2}, the preliminaries of our scheme are introduced. The system model, security requirements and design goals are described in Section \ref{sec3}. In Section \ref{sec4}, the proposed framework is presented in details. The security analysis and performance evaluation are discussed in Section \ref{sec5} and \ref{sec6}. Three applications based on the proposed privacy-preserving SVD framework are illustrated in Section \ref{sec7}. In Section \ref{sec8}, we discuss the related work, and finally conclude our current work in Section \ref{sec9}.

\section{Preliminaries}
\label{sec2}
In this section, the Paillier Cryptosystem \cite{paillier1999public}
%, bilinear pairing technique \cite{boneh2001identity} 
and Singular Value Decomposition \cite{golub2012matrix} which are the basis of the proposed framework are reviewed.
\subsection{Paillier Cryptosystem}
The Paillier Cryptosystem enables the addition and multiplication operations on plaintext through the specific linear algebraic manipulation conducted on the ciphertext. This property is extensively desired in many privacy-preserving applications \cite{sang2009privacy, zhong2007privacy, lu2012eppa}. In this paper, this feature allows fog devices to process the user data in encrypted form without leaking the data content. The Paillier Cryptosystem comprises three phases: key generation, encryption and decryption.
\begin{itemize}
	\item \emph{Key Generation: }Given one security parameter $\kappa$, generate two large prime numbers $p$, $q$, where $|p|=|q| = \kappa$. Then compute the $\mathcal{RSA}$ modulus $n  = p q$, $\lambda = lcm(p-1, q-1)$ and choose a generator $g \in \mathbb{Z}_{n^2}^{\ast}$. Define the function $L(u) = \frac{u - 1}{n}$ and calculate $\mu = (L(g^{\lambda} \bmod n^2))^{-1} \bmod n$. Then \emph{\textbf{PK}} $= (n, g)$ is published as the public key and \emph{\textbf{SK}} $= (\lambda, \mu)$ is kept as the corresponding private key.
	\item \emph{Encryption: }Given a message $m \in \mathbb{Z}_n$, randomly choose a number $r \in \mathbb{Z}_n^{\ast}$, the ciphertext could be calculated as $c = E(m, r) = g^m \cdot r^n \bmod n^2$.
	\item \emph{Decryption: }Given the ciphertext $c \in \mathbb{Z}_{n^2}^{\ast}$, the corresponding plaintext could be recovered as $m = D(c) = L(c^{\lambda}  \bmod n^2)\cdot \mu \bmod n$. Note that, the Paillier Cryptosystem is provably secure against chosen plaintext attack, and the correctness and security can be referred to \cite{paillier1999public}.
\end{itemize}
The homomorphic property of the Paillier Cryptosystem utilized in this work is: $E(m_1, r_1) \cdot E(m_2, r_2) = E(m_1+m_2, r_1\cdot r_2)$.
%\subsection{Bilinear Pairing}
%Let $\mathbb{G}$, $\mathbb{G}_T$ be two cyclic groups of the same prime order $q_1$, and $P$ be a generator of group $\mathbb{G}$. Suppose $\mathbb{G}$ and $\mathbb{G}_T$ are equipped with a pairing which is a nondegenerated and efficiently computable bilinear map $e: \mathbb{G} \times \mathbb{G} \rightarrow \mathbb{G}_T$ fulfilling: \emph{1)} there exits $P$ and $Q$ $\in \mathbb{G}$ such that $e(P, Q) \neq 1_{\mathbb{G}_T}$ \emph{2)} for all $a, b \in \mathbb{Z}_{q_1}^{\ast}$ and any $P_1, Q_1 \in \mathbb{G}$, there are $e(aP_1, bQ_1) = e(P_1, Q_1)^{ab} \in \mathbb{G}_T$. A more comprehensive description of pairing technique, and complexity assumptions could be found in \cite{boneh2001identity}.\\
%\textbf{Definition 1. }{\it The bilinear parameter generator $\mathcal{G}en$ is a probabilistic algorithm which takes a security parameter $\kappa_1$ as input and outputs a 5-tuple $(q_1, P, \mathbb{G}, \mathbb{G}_T, e)$ where $q_1$ is a prime number whose bit length is $\kappa_1$, $\mathbb{G}$, $\mathbb{G}_T$ are two groups with order $q_1$, $P$ is a generator of $\mathbb{G}$, and $e: \mathbb{G} \times \mathbb{G} \rightarrow \mathbb{G}_T$ is the bilinear map which is nondegenerated and efficiently computable.}
%
%The security of bilinear pairing could be derived from the hardness of solving the Computational Diffie-Hellman (CDH) Problem, Bilinear Diffie-Hellman (BDH) Problem or Decisional BDH (DBDH) Problem. We refer to \cite{boneh2001identity} for the detailed definition of CDH problem, BDH problem and DBDH problem.

\subsection{Singular Value Decomposition}
SVD is a powerful and popular matrix factorization tool that underlies plenty of useful applications, e.g. abnormal detection \cite{ide2004eigenspace, lee2013anomaly}, recommendation system \cite{sarwar2000application}\cite{billsus1998learning} and data compression \cite{kalman1996singularly}. Let $\mathbb{A}$ be a $l \times N$ matrix, the SVD of $\mathbb{A}$ is of the form  $\mathbb{U} \Sigma \mathbb{V}^{T}$ where $T$ means conjugate transpose, $\mathbb{U}$ is a $l \times l$ unitary matrix, $\Sigma$ is a $l \times N$ rectangular diagonal matrix with non-negative diagonal values, and $\mathbb{V}$ is an $N \times N$ unitary matrix. The non-negative diagonal entries of $\Sigma$ are the singular values of matrix $\mathbb{A}$ while the columns of $\mathbb{U}$ and $\mathbb{V}$ are known as the left-singular vectors and right-singular vectors of $\mathbb{A}$. Note that we only consider real matrix entries in this paper, thus the conjugate transpose $T$ could be simply regarded as transpose. 

Another widely used matrix factorization tool is the eigenvalue decomposition. Unlike SVD which could be applied to any $l \times N$ matrix, the eigenvalue decomposition is less general and could only be performed on square matrix. However, the two kinds of tool are closely related shown as follow:  
\begin{equation}
\begin{split}
\mathbb{A} \cdot \mathbb{A}^{T} = \mathbb{U}\Sigma \mathbb{V}^{T} \mathbb{V} {\Sigma}^{T} \mathbb{U}^{T} = \mathbb{U} \Sigma {\Sigma}^{T} \mathbb{U}^{T} \\
\mathbb{A}^{T} \cdot \mathbb{A} = \mathbb{V} {\Sigma}^{T} \mathbb{U}^{T}\mathbb{U} \Sigma \mathbb{V}^{T} = \mathbb{V} {\Sigma}^{T} \Sigma \mathbb{V}^{T}	
\end{split}
\end{equation}
It is obvious that the left-singular vectors $\mathbb{U}$ is the eigenvectors of $\mathbb{A} \cdot \mathbb{A}^{T}$, the right-singular vectors $\mathbb{V}$ is the eigenvectors of $\mathbb{A}^{T} \cdot \mathbb{A}$ and the singular values $\Sigma$ are the square root of the eigenvalues of $\mathbb{A} \cdot \mathbb{A}^{T}$ and $\mathbb{A}^{T} \cdot \mathbb{A}$. We will show that the above relation could be utilized to achieve the privacy-preserving SVD in the later sections.

\section{System Model, Security Requirements and Design Goals}
\label{sec3}

In this section, we describe the system model, discuss the security requirements and identify the design goals on privacy-preserving SVD.
\subsection{System Model}
\begin{figure}[htbp]
	\centering
	\includegraphics[width=0.5\textwidth]{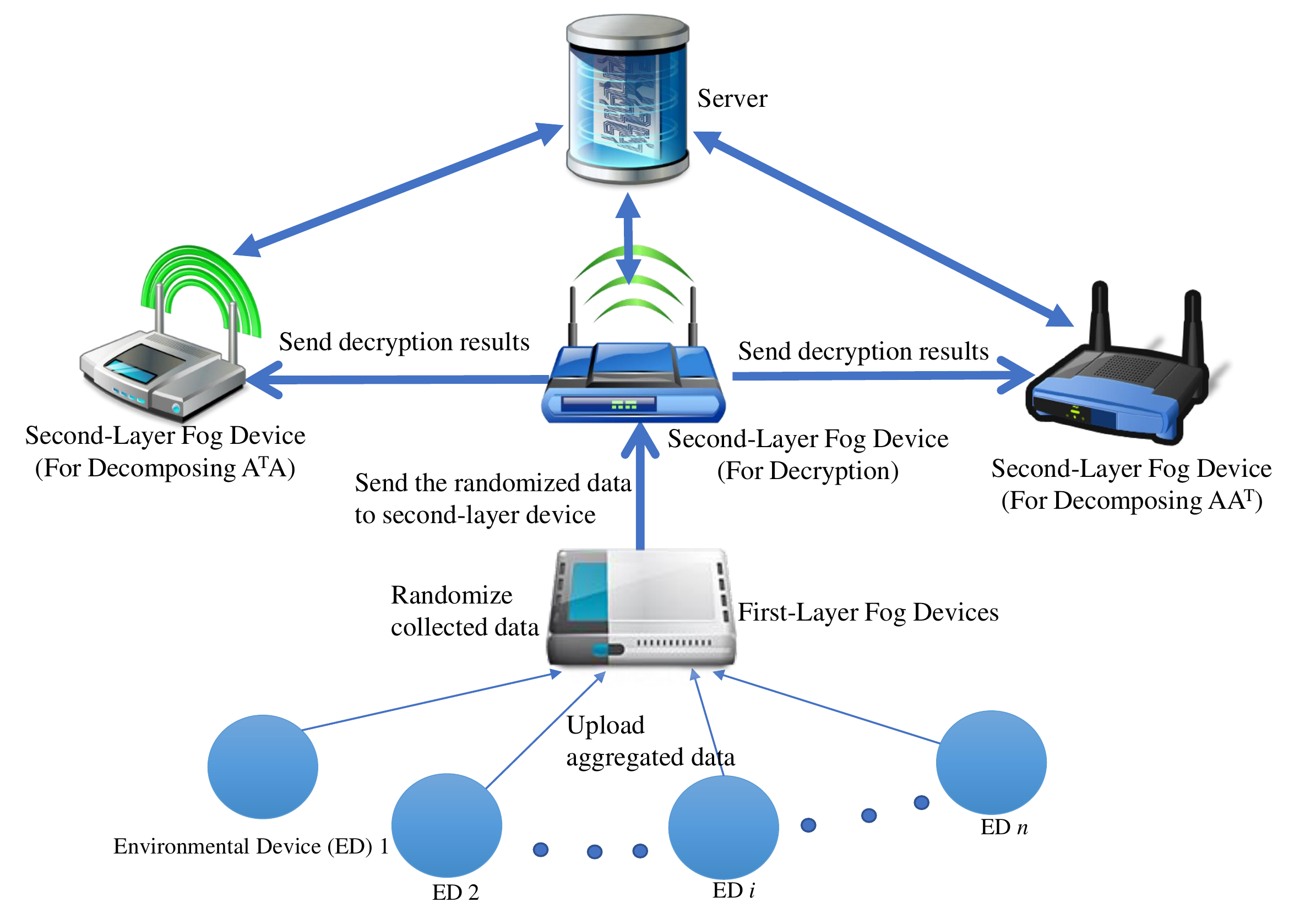}
	%\makebox[\textwidth]{\includegraphics[width=0.5\textwidth]{fig2}}\\
	\caption{System Model}\label{fig2}
\end{figure}

In this work, we mainly focus on how to utilize the fog computing to compute the SVD of the data uploaded by environmental devices with privacy preserved. Specifically, there are four categories of entity in the system model, namely server, first layer fog device, second layer fog device and environmental device as shown in Fig. \ref{fig2}.

\textbf{Server}: server is a fully trustable entity located in the central tier. It is responsible for initializing the whole system and distributing key materials to others. The other operations the server may conduct are application-specific. Serval examples will be given in Section \ref{sec7}.

\textbf{Environmental Device (ED)}: EDs are the devices distributed in the environmental layer of IoT environment. The analysis on the data uploaded by EDs could enable better decision-making.

\textbf{First Layer Fog Device (FD)}: FDs are the fog devices which communicate with EDs directly. FDs process the collected data and upload the results to the second layer fog devices. %Depending on the scale of local area, there could be only one FD or multiple FDs.

\textbf{Second Layer Fog Device (SD)}: SDs are the fog devices which communicate with FDs. Compared to FDs, SDs are closer to the server and do not contact with EDs directly. In the proposed framework, there are three SDs playing different roles for SVD operation. One of them is responsible for decrypting the messages from FDs. The other two are in charge of decomposing $\mathbb{A} \cdot \mathbb{A}^{T}$ and $\mathbb{A}^{T} \cdot \mathbb{A}$. We denote the one for decryption, decomposing $\mathbb{A} \cdot \mathbb{A}^{T}$ and decomposing $\mathbb{A}^{T} \cdot \mathbb{A}$ as SD$_d$, SD$_u$ and SD$_v$ respectively.

The basic workflow of our model is: the server first initializes the whole system and distributes the key materials or secrets to other entities accordingly. After initialization, EDs start to collect and upload application-specific data as required. Each data transmitted to FDs is encrypted. After FDs gather the data from EDs, FDs will randomize the collected data and send the results to SD$_d$. Upon decrypting the encrypted data from FDs, SD$_d$ will perform further operations on the plaintext and send the outcomes to SD$_u$ and SD$_v$ respectively. With the messages from SD$_d$, SD$_u$ and SD$_v$ could recover $\mathbb{A} \mathbb{A}^{T}$ and $\mathbb{A}^{T} \mathbb{A}$, and conduct the eigenvalue decomposition accordingly. Finally, the SVD of data collection is split into two parts which are held by SD$_u$ and SD$_v$ seperately.

\subsection{Security Requirements}
Security is fundamental for the effectiveness of proposed framework. In this work, the server and EDs are assumed to be trustable. The fog devices, i.e. FDs and SDs, are assumed to be honest-but-curious \cite{Goethals2004, Lindell2000} which means they will follow the specified procedures faithfully while being curious about the uploaded data. In addition, FDs and SDs are assumed not to collude with each other. The non-collusion assumption could be realized similarly as the EigenTrust scheme \cite{kamvar2003eigentrust}. Briefly speaking, for each SVD computation, the server chooses fog devices based on distributed hash table. Due to the large number of fog devices, it is infeasible for the device providers to determine whether they would be selected for the same computation and negotiate for collusion in advance. % \emph{2)} $\mathcal{A}$ could replay and modify the eavesdropped messages to launch some active attacks. 
Based on the above assumptions, the confidentiality as the security requirement should be fulfilled, i.e. even FDs and SDs process the collected data, they could not learn anything about the actual value of data. For authenticity and integrity, since there are many existing signature schemes for them, e.g. Boneh-Lynn-Shacham (BLS) short signature \cite{boneh2004short}, this work focuses on confidentiality. 

\subsection{Design Goals}
According to the aforementioned system model and security requirements, the goal is to design a flexible fog computing framework for privacy-preserving SVD computation. Specifically, the proposed framework should achieve the following objectives.
\begin{itemize}
	\item \emph{The confidentiality should be guaranteed in the proposed framework.} All the user data contained in the transmitted messages should be protected. The processing in fog devices should not leak data privacy.
	
	\item \emph{The framework should be flexible enough to be adopted by different applications.} Instead of being the ultimate goal, SVD is the basis or initial step of many applications, which means the further procedures after SVD could be quite different for various scenarios. Therefore, the design of the framework should consider the flexibility such that the results of SVD could be further utilized to achieve the final purposes of different applications.	 
\end{itemize}

\section{The Proposed Framework}
\label{sec4}
In this section, the proposed framework for SVD computation is presented in details. The framework is composed of five phases: system initialization, data collection, data randomization, pre-computation and eigenvalue decomposition.
\subsection{System Initialization}
The server is the trustable entity which bootstraps the whole system. Assume the amount of users supported by the system is $N$, each user data is $l$-dimensional and the range for each dimension value is [$0$, $d$] where $d$ is a constant. The system parameters are $\kappa$, $\kappa_1$, $\kappa_2$ and $\kappa_3$. Let $|\bullet|$ denote the bit length of $\bullet$. Given the parameter $\kappa$, the server calculates the public key for Paillier Cryptosystem \emph{\textbf{PK}}: $(n = p q, g)$, and the corresponding private key \emph{\textbf{SK}}: $(\lambda, \mu)$, where $p$, $q$ are two large primes with $|p| = |q| = \kappa$. Given the parameters $\kappa_1$, $\kappa_2$ and $\kappa_3$, let $t = 2^{\kappa_1}$, the server randomly chooses two coprime integers $W$ and $S$ such that $W > max(N, l)\cdot{d^2}$ and $S > max(N, l)\cdot{(d^2+2tWd+t^2W^2)}$, where $|W| = \kappa_2$ and $|S| = \kappa_3$. Then, the server chooses one superincreasing sequence $\vec{\mathbf{a}} = (a_1 = 1, a_2,\cdots, a_l)$ such that $\sum_{j = 1}^{i-1}a_j \cdot (d+tW+tS) < a_i$ for $i = 2,\cdots, l$ and  $\sum_{i=1}^{l}a_i \cdot (d+tW+tS) < n$. Finally, the server publishes $\{n, g, \vec{\mathbf{a}}\}$ as public parameters, sends $(\lambda, \mu)$ to SD$_d$ as secret, and sends $(W, S)$ to FDs, SD$_u$ and SD$_v$ as secret respectively.

\subsection{Data Collection}
In the environmental tier, EDs collect and upload the application-specific $l$-dimensional data $(d_1,\cdots, d_l)^T$. The data from $N$ EDs could form the data matrix $\mathbb{A}$.
$$
\mathbb{A} =
\begin{pmatrix}
d_{11}&d_{12}&\ldots&d_{1N}\\
d_{21}&d_{22}&\ldots&d_{2N}\\
\vdots &\vdots & \ddots &\vdots\\
d_{l1}&d_{l2}&\ldots&d_{lN}\\
\end{pmatrix}
$$
To compute the SVD of matrix $\mathbb{A}$ with privacy preserved is the goal of this work.
%To detect faulty sensors, we need the data samples being collected during the period when the patient's condition is stable. 
The $i$th column of matrix $\mathbb{A}$ $(d_{1i},\cdots, d_{li})^T$ is the data from the $i$th device ED$_i$. To upload the data, ED$_i$ performs the following steps:
\begin{itemize}
	\item \emph{Step-1. }Utilize the superincreasing $\vec{\mathbf{a}}$ to compute
	\begin{equation}
	\label{equSuper}
	m_i = a_1d_{1i}+a_2d_{2i}+\cdots+a_ld_{li}
	\end{equation}
	\item \emph{Step-2. }Choose a random number $r_i \in \mathbb{Z}_n^{\ast}$ and compute
	\begin{equation}
	C_i = g^{m_i} \cdot r_i^n \bmod n^2
	\end{equation}

	\item \emph{Step-3. }Send the encrypted data $C_i||ED_i$ to the FD which communicates with it.
\end{itemize}
\subsection{Data Randomization}
For each FD, it will perform the following steps to randomize the received data.
\begin{itemize}
	\item \emph{Step-1. }For the $i$th data $C_i$, FD chooses $2\cdot l$ random numbers which are  $(z_{1i},\cdots, z_{li})^T$ and  $(r_{1i},\cdots, r_{li})^T$ from the range $[1, t]$. Then FD computes $ rz = \sum_{k=1}^{l}a_k \cdot (z_{ki}\cdot W+r_{ki}\cdot S)$.
	
	\item \emph{Step-2. }FD randomizes $C_i$ as
	\begin{equation}
	\begin{split}
	{C_i}^{'} &= C_i \cdot g^{rz} \bmod n^2\\
	&= g^{\sum_{k=1}^{l}a_k \cdot d_{ki}} \cdot g^{\sum_{k=1}^{l}a_k \cdot (z_{ki}\cdot W+r_{ki}\cdot S)} \cdot r_i^n \bmod n^2\\
	&= g^{\sum_{k=1}^{l} a_k \cdot (d_{ki}+z_{ki}\cdot W+r_{ki}\cdot S)} \cdot r_i^n \bmod n^2\\
	%&= g^{a_1\sum_{i=1}^{N}d_{1i}+\cdots+a_l\sum_{i=1}^{N}d_{li}} \cdot \prod_{i=1}^{N}r_i^n \bmod n^2\\
	\end{split}
	\end{equation}
	\item \emph{Step-3. }FD sends the randomized data ${C_i}^{'}$ to SD$_d$.
\end{itemize}
\subsection{Pre-computation}
Upon receiving $N$ data from FDs, SD$_d$ will perform the following steps to compute the randomized $\mathbb{A} \mathbb{A}^{T}$ and $\mathbb{A}^{T} \mathbb{A}$. 

\begin{itemize}
	\item \emph{Step-1. }For each ${C_i}^{'}$, SD$_d$ will utilize the secret key $(\lambda, \mu)$ to decrypt it and get the aggregation of randomized data
	\begin{equation}
	{m_i}^{'} = \sum_{k=1}^{l}a_k \cdot (d_{ki}+z_{ki}\cdot W+r_{ki}\cdot S) \bmod n
	\end{equation} 
	\item \emph{Step-2. }Through running the Algorithm \ref{Alg:Cal}, SD$_d$ could recover the randomized value for each dimension of data $i$. %The correctness of the \textbf{Algorithm 1} could also be found in \cite{lu2012eppa}.
	\begin{algorithm}
		\footnotesize
		\caption{Recover randomized value from aggregated data}\label{Alg:Cal}
		\begin{algorithmic}[1]
			\Procedure {Recover randomized value of each dimension}{}
			
			\noindent\textbf{Input:} $\vec{\mathbf{a}}=(a_1=1, \cdots, a_l)$ and ${m_i}^{'}$ \newline
			\noindent\textbf{Output:} Randomized data
			\State{Let $\vec{Temp_a} = (t_{a1}, \cdots, t_{al})^T$}		
			\State{Set $X_l = {m_i}^{'}$}
			\For{$k = l$ to $2$}
			\State{$X_{k-1} = X_k \bmod a_k$}
			\State{$t_{ak} = \frac{X_k - X_{k-1}}{a_k} = d_{ki}+z_{ki}\cdot W+r_{ki}\cdot S$}
			
			\EndFor
			\State{$t_{a1} = X_1  = d_{1i}+z_{1i}\cdot W+r_{1i}\cdot S$}
			
			\State{\textbf{return} $(t_{a1}, \cdots, t_{al})^T$}
			\EndProcedure
		\end{algorithmic}
	\end{algorithm}
	
	\item \emph{Step-3. }From each ${m_i}^{'}$, SD$_d$ could get an $l$-dimensional randomized data. In total, SD$_d$ could get the randomized data matrix
	$$
	\mathbb{A}^{'} =
	\begin{pmatrix}
	d_{11}^{'}&d_{12}^{'}&\ldots&d_{1N}^{'}\\
	d_{21}^{'}&d_{22}^{'}&\ldots&d_{2N}^{'}\\
	\vdots &\vdots & \ddots &\vdots\\
	d_{l1}^{'}&d_{l2}^{'}&\ldots&d_{lN}^{'}\\
	\end{pmatrix}
	$$
The $(i, j)$th entry of $\mathbb{A}^{'}$ is $d_{ij}^{'} = d_{ij}+z_{ij}\cdot W+r_{ij}\cdot S$. Then SD$_d$ simply computes $\mathbb{A}^{'} \cdot (\mathbb{A}^{'})^{T}$ and  $(\mathbb{A}^{'})^{T} \cdot \mathbb{A}^{'}$, and sends the two resulting matrices to SD$_u$ and SD$_v$ respectively.
\end{itemize}

\emph{\textbf{The correctness of Algorithm~\ref{Alg:Cal}}}. In Algorithm~\ref{Alg:Cal}, $X_l = {m_i}^{'} = a_1 \cdot (d_{1i}+z_{1i}\cdot W+r_{1i}\cdot S)  + \cdots + a_{l-1} \cdot (d_{(l-1)i}+z_{(l-1)i}\cdot W+r_{(l-1)i}\cdot S) + a_l \cdot (d_{li}+z_{li}\cdot W+r_{li}\cdot S)] \bmod n$. Since the data value for each dimension is in the range of [$0$, $d$], and $z_{ki}$ and $r_{ki}$ are chosen from $[1, t]$, we have
\begin{equation}\label{}
\begin{split}
&a_1 \cdot (d_{1i}+z_{1i}\cdot W+r_{1i}\cdot S)  + \cdots + a_{l-1} \cdot (d_{(l-1)i}+z_{(l-1)i}\cdot W+r_{(l-1)i}\cdot S)\\
<& a_1 \cdot (d + tW + tS)+ \cdots + a_{l-1} \cdot(d + tW + tS) \\
=& \sum_{k=1}^{l-1} a_k \cdot(d + tW + tS) < a_l \\
\end{split}
\end{equation}
Therefore, $X_{l-1} = X_l \bmod a_l = a_1 \cdot (d_{1i}+z_{1i}\cdot W+r_{1i}\cdot S)  + \cdots + a_{l-1} \cdot (d_{(l-1)i}+z_{(l-1)i}\cdot W+r_{(l-1)i}\cdot S)$, and
\begin{equation}
\begin{split}
\frac{X_{l} - X_{l-1}}{a_l} &= \frac{a_l \cdot (d_{li}+z_{li}\cdot W+r_{li}\cdot S)}{a_l}\\ 
&= d_{li}+z_{li}\cdot W+r_{li}\cdot S\\
\end{split}
\end{equation}
Similarly, it can be proved that  $t_{ak} = d_{ki}+z_{ki}\cdot W+r_{ki}\cdot S$, for $k = 1, \cdots, l-1$. 

\subsection{Eigenvalue Decomposition}
\subsubsection{SD$_u$:}
When receiving $\mathbb{A}^{'} \cdot (\mathbb{A}^{'})^{T}$, SD$_u$ will perform the following steps to compute the left part of the SVD for matrix $\mathbb{A}$, i.e. matrix $\mathbb{U}$ and $\Sigma$.
\begin{itemize}
	\item \emph{Step-1. }For each entry ${e_u}^{'}$ of $\mathbb{A}^{'} \cdot (\mathbb{A}^{'})^{T}$, SD$_u$ derandomizes the entry as follow:
	\begin{equation}
	e_u = {e_u}^{'} \bmod S \bmod W
	\end{equation}
	The result $e_u$ is the corresponding entry of matrix  $\mathbb{A}\cdot \mathbb{A}^{T}$.
	\item \emph{Step-2. }After SD$_u$ recovers the matrix $\mathbb{A}\cdot \mathbb{A}^{T}$, it performs eigenvalue decomposition for matrix $\mathbb{A}\cdot \mathbb{A}^{T}$ and gets the matrix $\mathbb{U}$ and $\Sigma$.
\end{itemize}

\subsubsection{SD$_v$:}
Similar as SD$_u$, SD$_v$ will derandomize the entries of received $(\mathbb{A}^{'})^{T} \cdot \mathbb{A}^{'}$ to recover the matrix $\mathbb{A}^{T} \cdot \mathbb{A}$. Then SD$_v$ performs the eigenvalue decomposition on $\mathbb{A}^{T} \cdot \mathbb{A}$ to get the right part of $\mathbb{A}$'s SVD, i.e. matrix $\mathbb{V}$ and $\Sigma$.

By now, the SVD of $\mathbb{A}$ has been seperately held by SD$_u$ and SD$_v$.
\\

\emph{\textbf{The correctness of derandomization}} The $(i, j)$th entry ${e_u}^{'}_{ij}$ of $\mathbb{A}^{'} \cdot (\mathbb{A}^{'})^{T}$ is implicitly formed as
\begin{equation}
\label{formation9}
\begin{split}
{e_u}^{'}_{ij} = \sum_{k=1}^{N} d_{ik}^{'} \cdot d_{jk}^{'}
&= \sum_{k=1}^{N} (d_{ik}+z_{ik}\cdot W+r_{ik}\cdot S) \cdot (d_{jk}+z_{jk}\cdot W+r_{jk}\cdot S) \\
&= \sum_{k=1}^{N} d_{ik} \cdot d_{jk} + \sum_{k=1}^{N} [(z_{ik} d_{jk}+z_{jk} d_{ik})W+z_{ik}z_{jk}W^2] \\
&+ S\sum_{k=1}^{N}[(r_{ik} d_{jk}+r_{jk} d_{ik}) + (z_{ik} r_{jk}+z_{jk} r_{ik})W+r_{ik}r_{jk}S]\\
\end{split}
\end{equation}
Since 
%$\sum_{k=1}^{N} d_{ik} \cdot d_{jk} + \sum_{k=1}^{N} [(z_{ik}\cdot d_{jk}+z_{jk}\cdot d_{ik}) \cdot W +z_{ik}z_{jk}W^2] < N\cdot(d^2+2tWd+t^2W^2) < S$,
\begin{eqnarray*}
	%\begin{split}
			&\sum_{k=1}^{N} d_{ik} d_{jk} + \sum_{k=1}^{N} [(z_{ik} d_{jk}+z_{jk} d_{ik}) W +z_{ik}z_{jk}W^2] < N(d^2+2tdW+t^2W^2) < S,
	%\end{split}
\end{eqnarray*}
we have ${e_u}^{'}_{ij} \bmod S = \sum_{k=1}^{N} d_{ik} d_{jk} + \sum_{k=1}^{N} [(z_{ik} d_{jk}+z_{jk} d_{ik}) W +z_{ik}z_{jk}W^2]$. Also, we have $\sum_{k=1}^{N} d_{ik} d_{jk} < N d^2 <W$. Thus, $({e_u}^{'}_{ij} \bmod S) \bmod W = \sum_{k=1}^{N} d_{ik} d_{jk}$ which is the $(i, j)$th entry of matrix $\mathbb{A}\cdot \mathbb{A}^{T}$. 

Similarly, for the $(i, j)$th entry ${e_v}^{'}_{ij}$ of matrix $(\mathbb{A}^{'})^{T} \cdot \mathbb{A}^{'}$, 
%is implicitly formed as
%\begin{equation}
%\begin{split}
%{e_v}^{'}_{ij} &= \sum_{k=1}^{l} d_{ki}^{'} \cdot d_{kj}^{'}\\
%&= \sum_{k=1}^{l} d_{ki} \cdot d_{kj} + \sum_{k=1}^{l} [(z_{ki} d_{kj}+z_{kj} d_{ki}) W +z_{ki}z_{kj}W^2] \\
%&+ S\sum_{k=1}^{l}[(r_{ki} d_{kj}+r_{kj} d_{ki}) + (z_{ki} r_{kj}+z_{kj} r_{ki}) W+r_{ki}r_{kj}S]\\
%\end{split}
%\end{equation}
%Since $W > max(N, l)\cdot{d^2}$ and $S > max(N, l)\cdot(d^2+2tWd$\newline $+t^2W^2)$, 
we have $({e_v}^{'}_{ij} \bmod S) \bmod W = \sum_{k=1}^{l} d_{ki} \cdot d_{kj}$ which is the $(i, j)$th entry of matrix $\mathbb{A}^{T} \cdot \mathbb{A}$.

\section{Security Analysis}
\label{sec5}
In this section, the security properties of the proposed framework are analysed. As mentioned in the security requirements of Section \ref{sec3}, the participants of the framework will faithfully follow the defined work procedures while being curious about the user data. Thus, we first analyze the ability of each participant to learn private data under normal operations, i.e. the probability of leaking privacy when following legal processes. Then, the possible extra operations, denoted as potential attacks, which could be conducted by certain participants to snoop data are analyzed. The resistance of the framework against those attacks is discussed and the principles for system configuration are demonstrated.  

\subsection{Privacy Leakage Probability under Normal Operations}
\subsubsection{EDs:}
In the proposed framework, each ED will encrypt its data with Paillier Cryptosystem before uploading. Thus, each ED could only know the plaintext of its own message and learn nothing about the other EDs' data.

\subsubsection{FDs:}
The messages collected by the FDs are all Paillier ciphertext. Since the FDs do not have the private key for decrypting messages and it is assumed that there is no collusion among fog devices, the user data privacy is protected by the Paillier Cryptosystem no matter what operations are performed on the ciphertext by FDs.

\subsubsection{SD$_{d}$:}
SD$_d$ is the only fog device which has the private key for decryption. However, the plaintexts SD$_d$ could get are the randomized data, i.e. ${d_{ij}}^{'} = d_{ij}+z_{ij}W+r_{ij}S,$ for $i = 1, \cdots, l$ and $j = 1, \cdots, N$. Since SD$_d$ has no idea about the value of $z_{ij}, r_{ij}, W$ and $S$, it could not gain any knowledge of the original data.

\subsubsection{SD$_{u}$:}
Through derandomizing $\mathbb{A}^{'} \cdot (\mathbb{A}^{'})^{T}$, SD$_u$ obtains $\mathbb{A} \cdot \mathbb{A}^{T}$ and further gets $\mathbb{U}$ and $\Sigma$. Since $\mathbb{A} = \mathbb{U} \Sigma \mathbb{V}^{T}$, SD$_u$ needs to find the correct unitary matrix $\mathbb{V}$ to determine $\mathbb{A}$. Since there are infinite unitary matrices, SD$_u$ could not recover original data matrix $\mathbb{A}$ with only $\mathbb{U}$ and $\Sigma$.

\subsubsection{SD$_{v}$:}
Similar as SD$_u$, SD$_v$ could not recover data matrix $\mathbb{A}$ with only $\mathbb{V}$ and $\Sigma$.

%%\subsubsection{Adversary $\mathcal{A}$}
%The adversary $\mathcal{A}$ could get each of the encrypted user data $C_i$, $\mathbb{A}^{'} \cdot (\mathbb{A}^{'})^{T}$ and $(\mathbb{A}^{'})^{T} \cdot \mathbb{A}^{'}$ through eavesdropping. For encrypted data, since $\mathcal{A}$ does not have the private key, there is not much it could do on the ciphertext. From $\mathbb{A}^{'} \cdot (\mathbb{A}^{'})^{T}$ and $(\mathbb{A}^{'})^{T} \cdot \mathbb{A}^{'}$, it is easy for $\mathcal{A}$ to recover the randomized data matrix $\mathbb{A}^{'}$. However, since the value of $S$, $W$, each $r_{ij}$ and $z_{ij}$ are never transmitted in public, $\mathcal{A}$ could not derandomized the data.

Based on the above analysis, the data privacy is totally preserved in the proposed framework when the participants follow the defined procedures. In the following, the possible extra computations performed by participants to discover private data are considered.

\subsection{Potential Attacks}
For EDs and FDs, since they only have encrypted user data, they could not get any meaningful information no matter what operations they perform on the ciphertext. Therefore, we mainly discuss the potential attacks from SD$_d$, SD$_u$ and SD$_v$ in this part.

\noindent$\bullet$ \textbf{SD$_d$}:
%\subsubsection{SD$_d$:}
As mentioned above, the information SD$_d$ gets is the randomized data ${d_{ij}}^{'} = d_{ij}+z_{ij}W+r_{ij}S,$ for $i = 1, \cdots, l$ and $j = 1, \cdots, N$. What SD$_d$ needs to do is to find the value of $S$ and $W$ and recover the original data as
\begin{equation}
	d_{ij} = {d_{ij}}^{'} \bmod S \bmod W
\end{equation}
Since ${d_{ij}}$ is mixed with the random combination of $S$ and $W$, it is infeasible for SD$_d$ to determine $S$ and $W$ without additional information. Therefore, we consider the situations in which SD$_d$ knows some of the user data. With the knowledge of user data, the possible operations SD$_d$ could do are as follows:

\begin{itemize}
	\item \emph{Step-1.} For each known data, SD$_d$ converts the corresponding randomized data to the form $zW+rS$ by computing ${d}^{'}-d$. Let $LC$ denote the set of converted data and $LC_i = z_iW+r_iS$ denote the $i$th element of $LC$.
	
	\item \emph{Step-2.} SD$_d$ performs the brute force attack, i.e. tries all possible $S$. For each try, SD$_d$ performs (modulo $S$) operation on each $LC_i$. Then SD$_d$ computes the greatest common divisor (GCD) of the resulting set. If the GCD is larger than 1, it is the value of $W$ and the currently selected $S$ is the correct $S$.
\end{itemize}

The rationale behind this attack is: the probability of $k$ randomly chosen integers being coprime is $\frac{1}{\varsigma(k)}$, where $\varsigma(x)$ is Riemann zeta function \cite{nymann1972probability}. When $k$ is large, the probability that they are not coprime is negligible. Thus, after the modulo operations on $LC$, only when the chosen $S$ is correct, the elements of the resulting set are of the form $zW$ and have a GCD larger than 1, which is $W$.

Note that, in some cases, SD$_d$ could still form a set, in which the elements are of the form $zW+rS$ with high probability, even it does not know any user data. For example, if the data matrix is sparse, most of the randomized data is already of the desired form. Another case is that data range is not large enough compared to the amount of data, SD$_d$ could compute $DV = d_{ij}^{'} - d_{i^{'}j^{'}}^{'}$ for all possible pairwise combinations $(d_{ij}^{'}, d_{i^{'}j^{'}}^{'})$ and some of the resulting $DV$s will be of the desired form. For those cases, SD$_d$ could perform the brute force attack.

\emph{\textbf{Parameter Selection}}. To resist the brute force attack in the possible cases, $|S|$, i.e. $\kappa_3$, should be at least equal to 80. Moreover, SD$_d$ could compute $LDV = LC_i - LC_j$ for all possible combinations $(LC_i, LC_j)$. If certain combinations have the same $zW$ inside, those resulting $LDV$s would be of the form $(r_i-r_j)S$. SD$_d$ could learn $S$ efficiently by computing the GCD of those $LDV$s even when $|S|\geq 80$. To avoid the case, the randomly chosen $z_{ij}$ should be different with each other with high probability. The $z_{ij}$ is chosen from the range $[1, t]$, and the total number of $z_{ij}$ is $l\cdot N$. According to the generalized birthday problem \cite{wagner2002generalized}, the probability of at least two chosen $z_{ij}$ match is $1- \exp^{\frac{-(lN)^2}{2t}}$. Thus, the probability of no match is $\exp^{\frac{-(lN)^2}{2t}}$ and the parameter $\kappa_1$ which determines $t$ could be selected accordingly. Note that if FDs could cooperatively choose the set of $z_{ij}$ such that there is no match, then the range $t$ only needs to be larger than $l\cdot N$.
%
%Moreover, as mentioned in Section \ref{sec4}, the $W$ must be coprime with $S$. This

\noindent$\bullet$ \textbf{SD$_u$}:
SD$_u$ could get $\mathbb{U}$, $\Sigma$ and $\mathbb{U}\Sigma$:

$$
\mathbb{U}=
\begin{pmatrix}
u_{11}&u_{12}&\ldots&u_{1l}\\
u_{21}&u_{22}&\ldots&u_{2l}\\
\vdots &\vdots & \ddots &\vdots\\
u_{l1}&u_{l2}&\ldots&u_{ll}\\
\end{pmatrix}
$$
$$
\Sigma=
\begin{pmatrix}
\sigma_1&0&\ldots&0&0&0\\
0&\sigma_2&\ldots&0&0&0\\
\vdots &\vdots & \ddots &\vdots&\vdots&\vdots\\
0&0&\ldots&\sigma_{\lambda}&0&0\\
\vdots &\vdots &  &\vdots&\vdots&\vdots\\
0&0&\ldots&0&0&0\\
\end{pmatrix}
$$
$$
\mathbb{U}\Sigma=
\begin{pmatrix}
\sigma_1 u_{11}&\sigma_2 u_{12}&\ldots&\sigma_{\lambda}u_{1\lambda}&0&0\\
\sigma_1 u_{21}&\sigma_2 u_{22}&\ldots&\sigma_{\lambda}u_{2\lambda}&0&0\\
\vdots &\vdots & \ddots &\vdots&\vdots&\vdots\\
\sigma_1 u_{l1}&\sigma_2 u_{l2}&\ldots&\sigma_{\lambda}u_{l\lambda}&0&0\\
\end{pmatrix}
$$
where $\lambda$ is the rank of $\mathbb{A}$. The purpose of SD$_u$ is to find the matrix $\mathbb{V}$:
$$
\mathbb{V}=
\begin{pmatrix}
v_{11}&v_{12}&\ldots&v_{1N}\\
v_{21}&v_{22}&\ldots&v_{2N}\\
\vdots &\vdots & \ddots &\vdots\\
v_{N1}&v_{N2}&\ldots&v_{NN}\\
\end{pmatrix}
$$

Note that the first $\lambda$ elements of each row in $\mathbb{V}$ correspond to a column of $\mathbb{A}$. If SD$_u$ knows the left $\lambda$ columns of $\mathbb{V}$, it could recover the original data. Since there are infinite unitary matrices, SD$_u$ could not determine the correct $\mathbb{V}$ if it has no additional information. Thus, we assume that SD$_u$ could get $N^{'}$ original data in some cases. Note that using one data, i.e. one column of $\mathbb{A}$, could form $l$ equations for the same row of $\mathbb{V}$. Solving the equations from one user data, SD$_u$ could determine the first $\lambda$ elements of that row.

When $N^{'} < N-1$, since each row of $\mathbb{V}$ is linearly independent with each other, obtaining one row does not help to learn the other rows. Thus, SD$_u$ could not utilize the known data to learn the rest unknown data. 

When $N^{'} = N-1$, SD$_u$ could determine the first $\lambda$ elements of $(N-1)$ rows of $\mathbb{V}$, then the first $\lambda$ elements of the last unknown row could be determined due to $\sum_{i=1}^{N}{v_{ij}^2} = 1$. The last unknown user data could be recovered accordingly.

%When $\lambda = N-1$ or $N$, the case is a little different because $\mathbb{V}$ needs to be an unitary matrix. Specifically, each data could determine the first $\lambda$ elements of one row in $\mathbb{V}$. When $\lambda = N-1$, the last element of that row could be determined since $\sum_{j=1}^{N}{v_{ij}^2} = 1$. When $\lambda = N$, obviously the entire row could be determined directly by solving equations. Therefore, when SD$_u$ knows the data from $(N-1)$ users, it could determine $(N-1)$ rows of $V$. The $N$ rows of $\mathbb{V}$ could be regarded as the coordinate axis of $N$-dimensional space whose rotation degree of freedom is $N-1$. Thus, when SD$_u$ determines $(N-1)$ rows, the rotation degree of freedom of the coordinate axis reduces to 0, i.e. the last unknown row could also be determined. With the last row, SD$_u$ could recover the only unknown user data. In a word, when $\lambda = N-1$ or $N$, SD$_u$ could learn the unknown user data if it has the data of the other $(N-1)$ users.

\noindent$\bullet$ \textbf{SD$_v$}:
%\subsubsection{SD$_v$}
Similar as SD$_u$, the matrix which could be utilized by SD$_v$ is $\Sigma \mathbb{V}^{T}$:

%could get three parts through modulo operations: $\mathbb{P}_1 = \mathbb{A}^{T}\mathbb{A}$, $\mathbb{P}_2 = \mathbb{Z}^{T}\mathbb{A}+\mathbb{A}^{T}\mathbb{Z}+ W\mathbb{Z}^{T}\mathbb{Z}$ and $\mathbb{P}_3 = (\mathbb{A}^{T}+W\mathbb{Z}^{T})\mathbb{R}+\mathbb{R}^{T}(\mathbb{A}+W\mathbb{Z})+S\mathbb{R}^{T}\mathbb{R}$. Since SD$_v$ also does not have $\mathbb{R}$ and $\mathbb{Z}$, $\mathbb{P}_2$ and $\mathbb{P}_3$ could not provide any information for it. From $\mathbb{P}_1$, SD$_v$ could get $\Sigma, \mathbb{V}$ and 

%$$
%\mathbb{V}=
%\begin{pmatrix}
%v_{11}&v_{12}&\ldots&v_{1N}\\
%v_{21}&v_{22}&\ldots&v_{2N}\\
%\vdots &\vdots & \ddots &\vdots\\
%v_{N1}&v_{N2}&\ldots&v_{NN}\\
%\end{pmatrix}
%$$
%$$
%\Sigma=
%\begin{pmatrix}
%\sigma_1&0&\ldots&0&0&0\\
%0&\sigma_2&\ldots&0&0&0\\
%\vdots &\vdots & \ddots &\vdots&\vdots&\vdots\\
%0&0&\ldots&\sigma_{\lambda}&0&0\\
%\vdots &\vdots &  &\vdots&\vdots&\vdots\\
%0&0&\ldots&0&0&0\\
%\end{pmatrix}
%$$
$$
\Sigma\mathbb{V}^{T}=
\begin{pmatrix}
\sigma_1 v_{11}&\sigma_1 v_{21}&\ldots&\sigma_1v_{N1}\\
\sigma_2 v_{12}&\sigma_2 v_{22}&\ldots&\sigma_2v_{N2}\\
\vdots &\vdots & \ddots &\vdots\\
\sigma_{\lambda} v_{1\lambda}&\sigma_{\lambda} v_{2\lambda}&\ldots&\sigma_{\lambda}v_{N\lambda}\\
0&0&\ldots&0\\
\vdots&\vdots&\ddots&\vdots\\
0&0&\ldots&0\\
\end{pmatrix}
$$
where $\lambda$ is the rank of $\mathbb{A}$. The purpose of SD$_v$ is to find the matrix $\mathbb{U}$:
$$
\mathbb{U}=
\begin{pmatrix}
u_{11}&u_{12}&\ldots&u_{1l}\\
u_{21}&u_{22}&\ldots&u_{2l}\\
\vdots &\vdots & \ddots &\vdots\\
u_{l1}&u_{l2}&\ldots&u_{ll}\\
\end{pmatrix}
$$
Different from the case of SD$_u$, the first $\lambda$ elements of each row in $\mathbb{U}$ correspond to a row of $\mathbb{A}$, i.e. the data value of a certain dimension from all users. If SD$_v$ knows the left $\lambda$ columns of $\mathbb{U}$, it could recover the original data. Similarly, we assume that SD$_v$ could get $N^{'}$ linearly independent user data in some cases. We have "linearly independent" here because linearly dependent data would not produce new linearly independent equations. Thus, only the number of linearly independent data matters. Each user data $\vec{d}_i$, i.e. one column of $\mathbb{A}$, could form one equation for each row of $\mathbb{U}$:
\begin{equation}
\left\{
\begin{aligned}
\sigma_1 v_{i1}\cdot u_{11}+\cdots&+\sigma_{\lambda} v_{i\lambda} u_{1\lambda}= d_{1i}  \\
\vdots &\\
\sigma_1 v_{i1}\cdot u_{l1}+\cdots&+\sigma_{\lambda} v_{i\lambda} u_{l\lambda}= d_{li}  \\
\end{aligned}
\right.
\end{equation}

When $\lambda < l$, if  $N^{'} = \lambda$, SD$_v$ could form $\lambda$ linearly independent equations for each row of $\mathbb{U}$. Then through solving equations, SD$_v$ could determine the left $\lambda$ columns of $\mathbb{U}$ and thus recover the whole $\mathbb{A}$ which contains the other unknown user data. On the other hand, if $N^{'} < \lambda$, SD$_v$ could not recover the other unknown linearly independent data due to the lack of enough linearly independent equations. However, for the data which is linearly dependent with the known data, SD$_v$ could recover them because the columns of $\Sigma\mathbb{V}^{T}$ have the same linear relationship as those existing among user data.

When $\lambda = l$, there is an additional condition for solving the equations of $\mathbb{U}$, i.e. $\mathbb{U}$ is a unitary matrix. Specifically, the $l$ rows of $\mathbb{U}$ could be regarded as the coordinate axis of $l$-dimensional space whose rotation degree of freedom is $l-1$. For each linearly independent data known to SD$_v$, the rotation degree of freedom of the coordinate axis reduces by 1. Therefore, if $N^{'} = l-1$, the rotation degree of freedom of the coordinate axis reduces to 0, i.e. the coordinate axis is fixed. Moreover, since $\sum_{j=1}^{l}{u_{ij}^2} = 1$, each row of $\mathbb{U}$ could be seen as a point locating on the unit sphere of $l$-dimension. Thus, the intersection points between the fixed coordinate axis and the $l$-dimensional unit sphere is the solution of $\mathbb{U}$.

Based on the above analysis, the proposed framework could resist the potential attacks launched by SD$_d$ through properly choosing $z_{ij}$ and $S$. For SD$_u$, only when $(N-1)$ user data is obtained, it could learn the value of the last unknown data. For SD$_v$, if $\lambda < l$, the framework could resist not more than $(\lambda-1)$ linearly independent user data leakage and it could resist not more than $(\lambda-2)$ linearly independent user data leakage if $\lambda = l$. 
\section{Performance Evaluation}
\label{sec6}

%\begin{table}[htbp]
%	\centering
%	\caption{Notations for Evaluation}
%	\label{Notations}
%	\begin{tabular}{|p{0.05\textwidth}|p{0.28\textwidth}|}
%		\hline
%		$l$ & Dimension of data\\
%		\hline
%		$N$ & Required number of data for detection\\
%		\hline
%		$N_s$ & Amount of sensors in the charge of one BCU\\
%		\hline
%		
%	\end{tabular}
%	
%\end{table}

In this section, we evaluate the performance of the proposed fog computing framework in terms of the capacity and efficiency. The capacity demonstrates the number of required ciphertexts for different matrix sizes while the efficiency indicates the computational complexity and communication overhead of the framework.
%The notations used in analysis are shown as table \ref{Notations}.
\subsection{Capacity}
In the proposed framework, 
%before EDs upload their user data, they will utilize the superincreasing sequence to aggregate the data of different dimensions as shown in equation \ref{equSuper}. After that, FDs will randomize the data with multiplication on Paillier ciphertext. Finally,
the aggregated randomized data is of the form ${m_i}^{'} = \sum_{k=1}^{l}a_k \cdot (d_{ki}+z_{ki}\cdot W+r_{ki}\cdot S)$. To guarantee the aggregated data could be recovered correctly through decryption, ${m_i}^{'}$ should be less than $n$, i.e. the constraint $\sum_{k=1}^{l}a_k \cdot (d+tW+tS) < n$ must be fulfilled. At the same time, the superincreasing sequence $\vec{\mathbf{a}}$ needs to meet the constraint: $\sum_{j = 1}^{i-1}a_j \cdot (d+tW+tS) < a_i$ for $i = 2,\cdots, l$. Moreover, in order to successfully derandomize the data, $W$ and $S$ need to fulfill: $W > max(N, l)\cdot{d^2}$ and $S > max(N, l)\cdot{(d^2+2tWd+t^2W^2)}$. To resist the potential attacks from SD$_d$, the value of $t$ should be chosen based on $N$ and $l$, and $\kappa_3$ should not be less than 80.% Therefore, whether the aggregated randomized data could be recovered from one Paillier ciphertext is determined by the value of $N$, $l$ and $d$. If the range of $N$, $l$ and $d$ are too large, it is definitely not enough for one Paillier ciphertext to accomodate the aggregated data. 

Let $\kappa_N, \kappa_l$ and $\kappa_d$ denote the bit length of $N$, $l$ and $d$ respectively. To resist the attacks from SD$_d$, $\kappa_1$ should be larger than $\kappa_N+\kappa_l$. For simplicity, assume that FDs could cooperatively select the $z_{ij}$ such that no match happens. Then $\kappa_1 = \kappa_N+\kappa_l+1$ is enough. To meet $W > max(N, l)\cdot{d^2}$, we have $\kappa_2 > max(\kappa_N, \kappa_l)+2\kappa_d$. Then due to $S > max(N, l)\cdot{(d^2+2tWd+t^2W^2)}$, $\kappa_3 > max(\kappa_N, \kappa_l)+2\kappa_1+2\kappa_2 > 3\cdot max(\kappa_N, \kappa_l)+4\kappa_d+2\kappa_N+2\kappa_l+2$. For the superincreasing sequence $\vec{\mathbf{a}}$, $|a_2| > \kappa_3+\kappa_1$ and $|a_3| > 2 (\kappa_3+\kappa_1)$. It is easy to find that $|a_i| > (i-1)(\kappa_3+\kappa_1)$ and $|\sum_{k=1}^{l}a_k \cdot (d+tW+tS)|>l(\kappa_3+\kappa_1)$. Thus, the bit length of aggregated data:
\begingroup
\renewcommand*{\arraystretch}{0.5}
\begin{equation*}
	|{m_i}^{'}| = 
	\left\{
	\begin{array}{ll}
		l[3max(\kappa_N, \kappa_l)+4\kappa_d+3\kappa_N+3\kappa_l+3] &\mbox{, if } max(\kappa_N, \kappa_l)+2\kappa_1+2\kappa_2 > 80.  \\
		\\
		l(80+\kappa_N+\kappa_l+1) &\mbox{, else.} \\
	\end{array}
	\right.
\end{equation*}
\endgroup

\begin{figure*}[!htbp]
	\begin{center}
		\vspace{-1\baselineskip}
		\subfigure[$|n|$ = 1024]
		{\label{n2048}\includegraphics[width=0.48\textwidth]{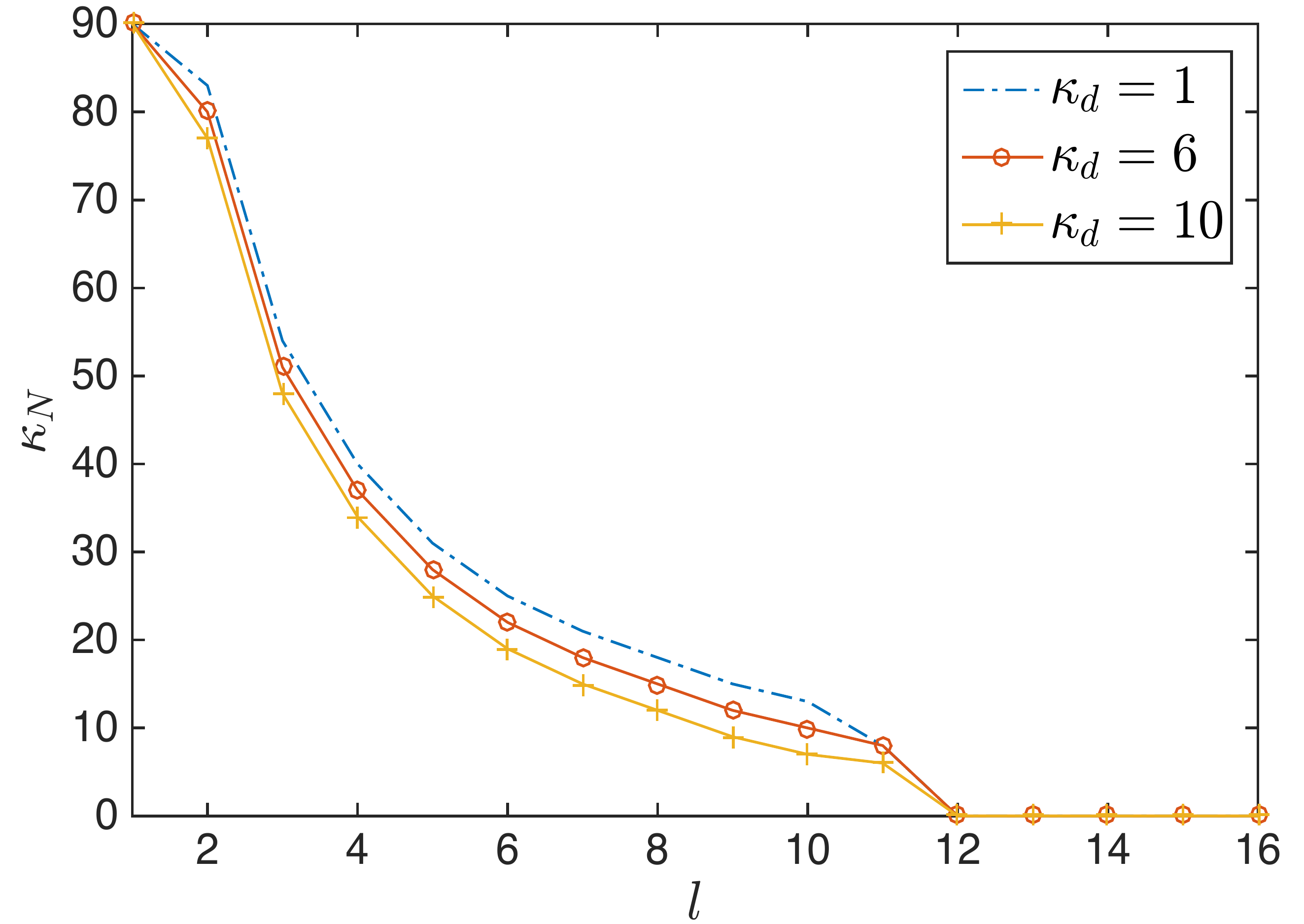}}
		%\hspace{-0.5cm}
		\subfigure[$\kappa_N$ = 10]
		{\label{n4096}\includegraphics[width=0.48\textwidth]{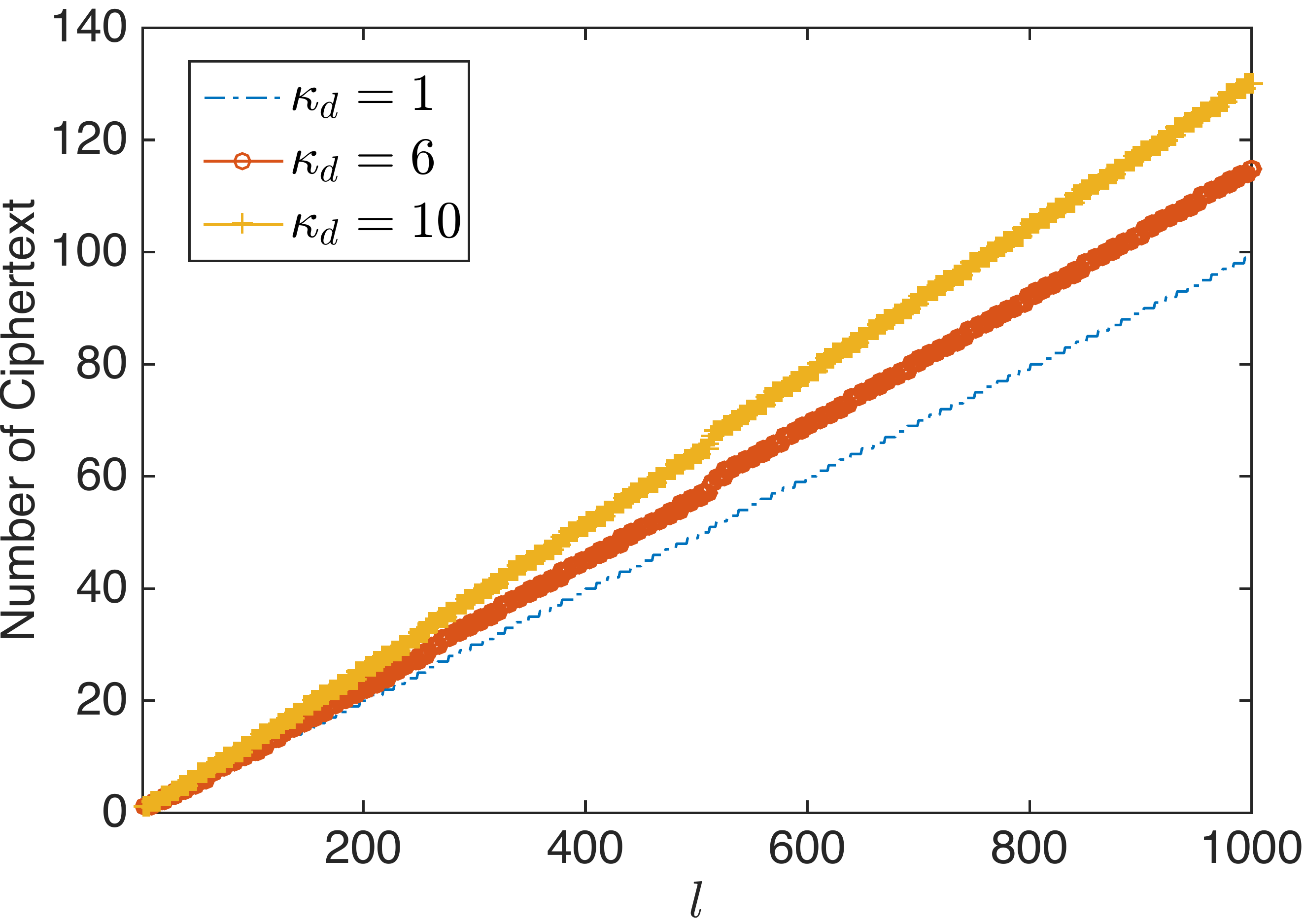}}
		\subfigure[$l$ = 150]	
		{\label{fig3_3}\includegraphics[width=0.48\textwidth]{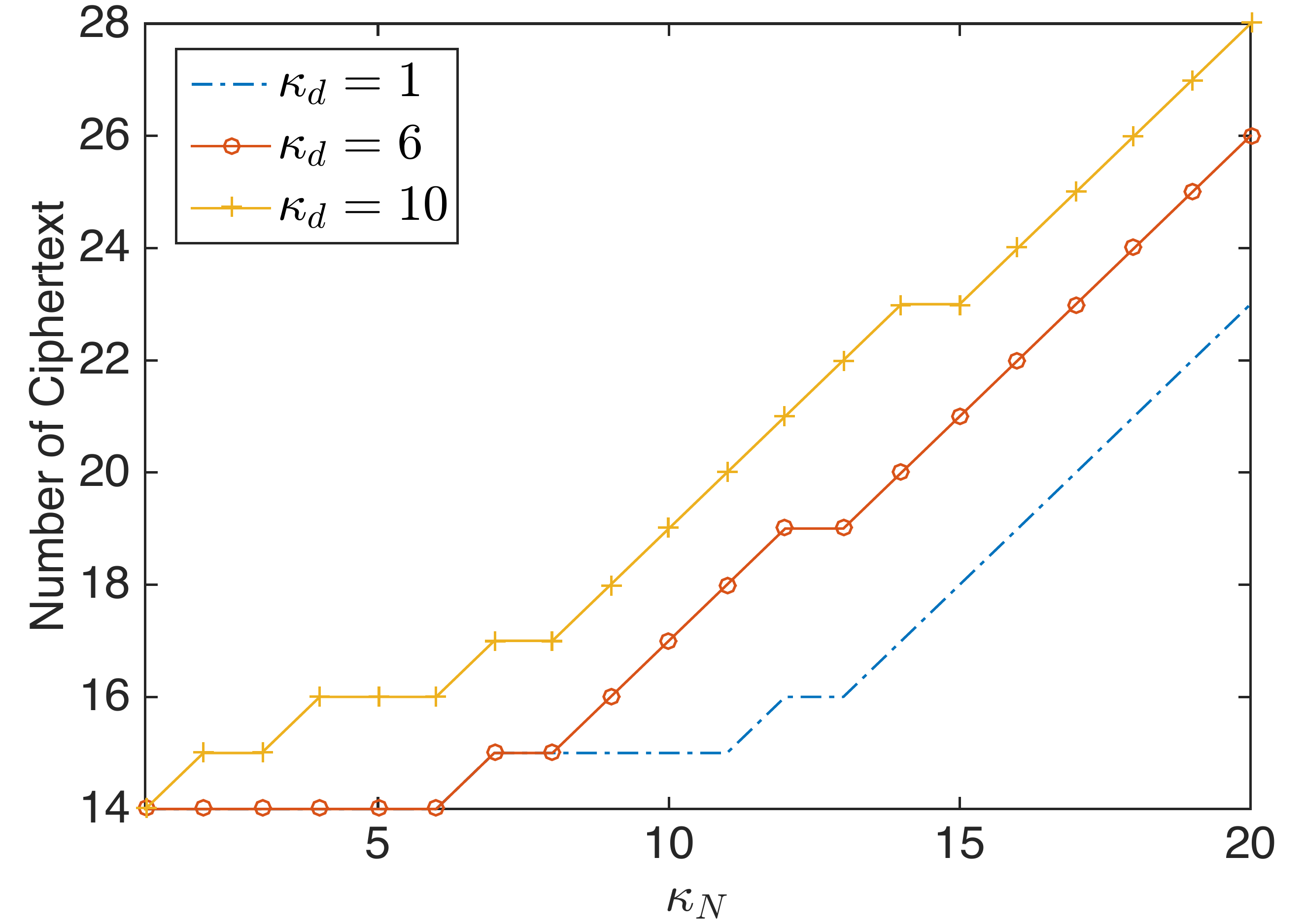}}
	\end{center}
	\vspace{-1\baselineskip}
	\caption{Capacity of the proposed framework}
	\vspace{-1\baselineskip}
	\label{fig3}
\end{figure*}

It is obvious that the data dimension has a great influence on the aggregated data length. Given different $\kappa_d$ and $l$, the number of users which one ciphertext with $|n| = 1024$ could support is evaluated as shown in Fig. \ref{n2048}.

From Fig. \ref{n2048}, it could be seen that the increase of dimensionality could dramatically decrease the number of users which one ciphertext could support, while the impact of data range $d$ is not that significant. One ciphertext could support large number of users with low dimensional data, e.g. $2^{37}$ users with $4$-dimensional data and $2^{15}$ users with $8$-dimensional data. To support higher dimensional data for the same amount of users, each ED needs to use multiple ciphertexts to aggregate data, e.g. to support $2^{15}$ users with 16-dimensional data needs 2 ciphertexts each of which aggregates 8 dimensions. Given different $l$ and $\kappa_N$, the number of required ciphertexts with $|n| = 1024$ is evaluated in Fig. \ref{n4096} and Fig. \ref{fig3_3} respectively. It could be seen that each ED needs to use $O(l\cdot log(N))$ ciphertexts for uploading data.

\subsection{Efficiency}
As analyzed above, each ED may need more than one Paillier ciphertext for aggregating user data. Let $N_C$ denote the number of required ciphertexts for each ED.
%could be computed as
%\begin{equation}
%N_C = \frac{|n|}{l\cdot[3\cdot max(\kappa_N, \kappa_l)+4\kappa_d+3\kappa_N+3\kappa_l+3]}
%\end{equation}
%where $|n|$ means the bit length of $n$. 
In the following, the computational complexity and communication overhead of the proposed framework are analyzed.

\noindent\textbf{Computational Complexity}: Since the crypto-operations are much heavier than the computations on plaintext, the amount of crypto-operations is the main concern in this part. The overall crypto-operations performed by each entity in the procedures of proposed framework are shown as follow.

$\bullet$ ED: each encryption needs $2$ exponentiation and $1$ multiplication in $\mathbb{Z}_{n^2}$. The overall crypto-operations conducted by an ED is $2\cdot N_C$ exponentiation and $N_C$ multiplication in $\mathbb{Z}_{n^2}$.

$\bullet$ FD: each randomization needs $1$ exponentiation and $1$ multiplication in $\mathbb{Z}_{n^2}$. Assuming the number of EDs communicating with the FD is $N_e$, the overall crypto-operations performed by a FD is $N_e\cdot N_C$ exponentiation and multiplication in $\mathbb{Z}_{n^2}$ respectively.

$\bullet$ SD$_d$: to decrypt one ciphertext, SD$_d$ needs to perform $1$ exponentiation operation in $\mathbb{Z}_{n^2}$. After decryption, the other operations are conducted on plaintexts and those cost is negligible compared to decryption. The overall crypto-operations conducted by the SD$_d$ is $N \cdot N_C$ exponentiation in $\mathbb{Z}_{n^2}$.

Note that SD$_u$ and SD$_v$ only perform computations on plaintext, and server is only in charge of system initialization. Therefore, their computation cost is negligible compared to the other entities.

Since the fog computing platform in current stage possesses the resource comparable to that of a smart phone, we have implemented the Paillier Cryptosystem on an Android mobile phone. The model number of the phone is Huawei Honor 3C (H30-U10) with the system parameters as: ARM Cortex-A7 4-core CPU @1.3GHz, 2GB memory and 4.2.2 Android version. When $|n|$ = 1024, the average running time (1000 iterations) for the exponentiation in $\mathbb{Z}_{n^2}$ is 55.493 milliseconds, the time for the multiplication in $\mathbb{Z}_{n^2}$ is 0.201 milliseconds and the time for the multiplication in $\mathbb{Z}_{n}$ is 0.101 milliseconds. It is obvious that the cost of multiplication is negligible compared to that of exponentiation. The computational cost for different entities is as shown in Table \ref{Computational_cost}.  
\begin{table}[htbp]
	\centering
	\caption{Computational Cost of the Proposed Framework}
	\label{Computational_cost}
	%\begin{tabular}{|p{0.4\textwidth}|p{0.4\textwidth}|}
	\begin{tabular}{|c|c|}
		\hline
		Entity & Computational Cost (milliseconds)\\
		\hline
		ED & $ 2 \times 55.493 \cdot N_C $\\
		\hline
		FDs & $55.493 \cdot N\cdot N_C$\\
		\hline
		SD$_d$ & $55.493 \cdot N\cdot N_C$\\
		\hline	
	\end{tabular}
\end{table}

Another notable thing is that the evaluation here implicitly assumes the IoT environment devices are as powerful as the smart phone. This is true for some IoT applications which use mobile phones or vehicles to upload environmental information. However, for the applications utilizing low power sensors as EDs, the Paillier operations are still too heavy. To circumvent this issue, the sensor may transmit its data to nearby more powerful device for conducting the crypto-operations. For example, the wristband  could connect with the mobile phone for processing and uploading data.

\noindent\textbf{Communication Overhead}:
%\subsubsection{Communication Overhead}
In this part, the communication overhead during SVD computation is evaluated. Note that for Paillier Cryptosystem, the ciphertext space is $\mathbb{Z}_{n^2}$. Thus, the bit length of one ciphertext is $2|n|$. The overhead of each communication flow is as shown in Table \ref{Communication_cost}.
\begin{table}[htbp]
	\centering
	\caption{Communication Overhead of the Proposed Framework}
	\label{Communication_cost}
	\centering
	\begin{tabular}{|c|c|}
		\hline
		Communication Flow & Bit Length of Message\\
		\hline
		ED$\rightarrow$ FD & $N_C \cdot 2|n|$\\
		\hline
		FDs $\rightarrow$ SD$_d$ & $N\cdot N_C \cdot 2|n|$\\
		\hline
		\rule{0pt}{0.3cm}
		SD$_d$ $\rightarrow$ SD$_u$ & $l^2(2\kappa_1+2\kappa_3+\kappa_N)$\\
		\hline
		\rule{0pt}{0.3cm}
		SD$_d$ $\rightarrow$ SD$_v$ & $N^2(2\kappa_1+2\kappa_3+\kappa_l)$\\
		\hline	
	\end{tabular}
\end{table}

For the "ED$\rightarrow$ FD" communication flow, each ED transits total $N_C$ ciphertexts to its corresponding FD, so the communication overhead is $N_C \cdot 2|n|$. For the "FDs $\rightarrow$ SD$_d$" communication flow, the overall ciphertexts SD$_d$ gathers from all the FDs is $N\cdot N_C$, so the communication overhead is $N\cdot N_C \cdot 2|n|$. For the "SD$_d$ $\rightarrow$ SD$_u$" communication flow, SD$_d$ sends $\mathbb{A}^{'} \cdot (\mathbb{A}^{'})^{T}$ to SD$_u$. Since the bit length of each entry in $\mathbb{A}^{'} \cdot (\mathbb{A}^{'})^{T}$ is $(2\kappa_1+2\kappa_3+\kappa_N)$ and there are total $l^2$ entries, the communication overhead is $l^2(2\kappa_1+2\kappa_3+\kappa_N)$. Similarly, the communication overhead for the "SD$_d$ $\rightarrow$ SD$_v$" communication flow is $N^2(2\kappa_1+2\kappa_3+\kappa_l)$.
\section{Applications}
\label{sec7}
In this section, we describe the potential IoT applications which could utilize the proposed framework. Basically, the proposed framework could be applied if the application possesses the following characteristics: 1) the application collects the environmental information for data analysis; 2) the data analysis is based on SVD; 3) the number of data analysis tasks is huge; 4) the environmental information is considered as privacy by the application users. Actually, the last two characteristics are the motivation of this work. The large amount of data analysis tasks motivates us to analyze data on fog computing platform. The privacy concern requires the analysis being privacy-preserving. Moreover, since different applications may utilize the result of SVD operation in different ways, we discuss how to adopt the proposed framework to achieve the purposes of different applications. In the following, three applications with different objectives are given as example to show that we only need to add several additional procedures or make some slight adjustments, the framework could be adapted to specific applications, which means the proposed framework is flexible.

\subsection{Abnormal Detection}
Since SVD could find the singular values of data matrix which reflect the features of data, it has proposed to apply SVD or the closely related technique--principal component analysis into abnormal detection \cite{ide2004eigenspace}\cite{lee2013anomaly}. The basic idea is quite straightforward: since the abnormal data would cause higher variation on the first eigenvector than the normal data, the variation degree of the first eigenvector could be utilized as the indicator of abnormal. When new data comes in, the system could perform SVD on the data matrix and check how much the direction of the first eigenvector changes. Note that to get the correct first eigenvector, it is usually required to normalize the data before performing SVD. However, the currently proposed framework is the general form which does not contain the normalization procedure. Thus, in the below, we show how to adjust the framework to accomplish that goal. Note that the eigenvector used for detection is from matrix $\mathbb{U}$, thus we only need to make adjustments on the procedures which are related to computing $\mathbb{U}$.

To include the normalization procedure, the $W$ and $S$ need to be chosen such that $W > 2{N^3d^2}$ and $S > 2{N^3(d^2+2tWd+t^2W^2)}$. Then, it needs to add some extra steps in the pre-computation operation of SD$_d$ and derandomization operation of SD$_u$. The extra steps are shown as follows.
\begin{itemize}
	\item\emph{Step-1. }For each randomized data $d_{ik}^{'}$, SD$_d$ performs
	\begin{equation}
	d_{ik}^{*}= N\cdot d_{ik}^{'}-\sum_{m=1}^{N}d_{im}^{'}	
	\end{equation}
	Then SD$_d$ performs matrix multiplication as before and sends the resulting $\mathbb{A}^{'}\cdot (\mathbb{A}^{'})^{T}$ to SD$_u$.
	\item\emph{Step-2. }For each entry $e$ of $\mathbb{A}^{'}\cdot (\mathbb{A}^{'})^{T}$, SD$_u$ computes $e^{'} = e \bmod S$ first. If $e^{'} > N^3(d^2+2tWd+t^2W^2)$, SD$_u$ further computes $e^{'} = e^{'}-S$. Then SD$_u$ computes $e^{''} = e^{'} \bmod W$. If $e^{''} > {N^3d^2}$, SD$_u$ further computes $e^{''} = e^{''}-W$. At last, SD$_u$ computes $e^{''} =\frac{e^{''}}{N^2}$. We denote the resulting matrix as $\mathbb{A}^{*}\cdot (\mathbb{A}^{*})^{T}$.
	\item\emph{Step-3. }Let $e^{*}_{ij}$ denote the $i$th row and $j$th column entry of $\mathbb{A}^{*}\cdot (\mathbb{A}^{*})^{T}$. SD$_u$ computes the standard deviation of the $i$th dimensional data as
	\begin{equation}
	\delta_{i} = \sqrt{\frac{e^{*}_{ii}}{N-1}}	
	\end{equation}
	Then for each entry $e^{*}_{ij}$, SD$_u$ computes $e^{*}_{ij} = \frac{e^{*}_{ij}}{\delta_{i} \delta_{j}}$. Finally, the resulting matrix is the correlation matrix generated from normalized data. The SD$_u$ could perform eigenvalue decomposition on the matrix as defined in the framework to find the first eigenvector and perform detection. One additional advantage of the proposed framework is that when new data arrives or the server wants to delete certain data, SD$_d$ only needs to add or substract the related randomized data, which is very convenient. 
\end{itemize} 

\emph{\textbf{The correctness of normalization}}. The $d_{ik}^{*}$ is implicitly formed by
\begin{equation}
\begin{split}
d_{ik}^{*} &= N\cdot d_{ik}^{'}-\sum_{m=1}^{N}d_{im}^{'}\\
&=(Nd_{ik}-\sum_{m=1}^{N}d_{im})+(Nz_{ik}-\sum_{m=1}^{N}z_{im})W+(Nr_{ik}-\sum_{m=1}^{N}r_{im})S\\
\end{split}
\end{equation}
Let $\hat{d_{ik}}$ denote $(Nd_{ik}-\sum_{m=1}^{N}d_{im})$, $\hat{z_{ik}}$ denote $(Nz_{ik}-\sum_{m=1}^{N}z_{im})$ and  $\hat{r_{ik}}$ denote $(Nr_{ik}-\sum_{m=1}^{N}r_{im})$. We have $-Nd<\hat{d_{ik}}<Nd$, $-Nt<\hat{z_{ik}}<Nt$ and $-Nt<\hat{r_{ik}}<Nt$. The $(i,j)$th entry of $\mathbb{A}^{'}\cdot (\mathbb{A}^{'})^{T}$ is implicitly formed by
\begin{equation}
\begin{split}
{e_{ij}} &= \sum_{k=1}^{N} d_{ik}^{*} \cdot d_{jk}^{*}\\
&= \sum_{k=1}^{N} \hat{d_{ik}} \cdot \hat{d_{jk}} + \sum_{k=1}^{N} [(\hat{z_{ik}} \hat{d_{jk}}+\hat{z_{jk}} \hat{d_{ik}}) W +\hat{z_{ik}}\hat{z_{jk}}W^2] \\
&+ S\sum_{k=1}^{N}[(\hat{r_{ik}} \hat{d_{jk}}+\hat{r_{jk}} \hat{d_{ik}}) + (\hat{z_{ik}} \hat{r_{jk}}+\hat{z_{jk}} \hat{r_{ik}})W +\hat{r_{ik}}\hat{r_{jk}}S]\\
\end{split}
\end{equation}
It is easy to infer that $-N^3d^2 < \sum_{k=1}^{N} \hat{d_{ik}} \cdot \hat{d_{jk}} < N^3d^2$ and $-N^3(d^2+2tWd+t^2W^2)<\sum_{k=1}^{N} \hat{d_{ik}} \cdot \hat{d_{jk}} + \sum_{k=1}^{N} [(\hat{z_{ik}}\cdot \hat{d_{jk}}+\hat{z_{jk}}\cdot \hat{d_{ik}}) \cdot W +\hat{z_{ik}}\hat{z_{jk}}W^2]<N^3(d^2+2tWd+t^2W^2)$. On the other hand, we have $S > 2{N^3(d^2+2tWd+t^2W^2)}$ and $W > 2{N^3d^2}$. Thus, if $e_{ij}^{'}$, which is $(e_{ij} \bmod S)$, is larger than $N^3(d^2+2tWd+t^2W^2)$, it means $\sum_{k=1}^{N} \hat{d_{ik}} \cdot \hat{d_{jk}} + \sum_{k=1}^{N} [(\hat{z_{ik}}\cdot \hat{d_{jk}}+\hat{z_{jk}}\cdot \hat{d_{ik}}) \cdot W +\hat{z_{ik}}\hat{z_{jk}}W^2] < 0$ and SD$_u$ needs to substract $S$ from the $e_{ij}^{'}$ to get the correct result. Similarly, when $e_{ij}^{''}$, which is $(e_{ij}^{'} \bmod W)$, is larger than $N^3d^2$, it means $\sum_{k=1}^{N} \hat{d_{ik}} \cdot \hat{d_{jk}} < 0$, SD$_u$ needs to substract $W$. Therefore, during \emph{Step-2}, SD$_u$ could recover the value of $e_{ij}^{''}=\sum_{k=1}^{N} \hat{d_{ik}} \cdot \hat{d_{jk}}$. It is obvious that $e_{ij}^{*}=\frac{e_{ij}^{''}}{N^2} = \sum_{k=1}^{N} (d_{ik}-\frac{1}{N}\sum_{m=1}^{N}d_{im})(d_{jk}-\frac{1}{N}\sum_{m=1}^{N}d_{jm})$ and $\sqrt{\frac{e_{ii}^{*}}{N-1} } =\sqrt{\frac{\sum_{k=1}^{N} (d_{ik}-\frac{1}{N}\sum_{m=1}^{N}d_{im})^2}{N-1}} = \delta_i$. The last step of dividing standard deviation is also straightforward.

\subsection{Localized Recommendation System}
SVD is also the underlying matrix factorization algorithm of Latent Semantic Indexing (LSI) which is widely used in information retrieval. Therefore, SVD is capable of capturing the latent
relationships among customers and products. In this part, we describe how to utilize the proposed framework to build up a recommendation system. %based on the scheme proposed in \cite{sarwar2000application}.

Imagine that there are tens of restaurants in the local region where you live. You have already been some of them and want to try a new one, let's say restaurant $p$, tonight. Before you go there, you would like to get a reputation score about $p$ from other people in the same region. If the score is too low, you may change the plan. We call the recommendation of local resource as localized recommendation. One may wonder why we consider a localized food recommendation system when there are already some centralized food review sites. The reason is that the localized food recommendation system has three advantages over the centralized food review sites. In specific, the first advantage is that the localized food recommendation system allows much more small restaurants to be included in the system while the centralized food review sites normally only have  the data of the restaurants of middle or larger size. For example, the canteens in a university campus or even the food windows inside a canteen could be recorded in the localized food recommendation system while it is hardly possible for the centralized food review sites to maintain the information of the vast amount of tiny food stalls. The second advantage is that the rating vectors uploaded by the reviewers in the localized food recommendation system could often provide meaningful recommendations to the query user because those people live in the same local region and they are likely to have been in the same restaurants. Thus, the size of data matrix required in localized recommendation system is much smaller. The third advantage is that the localized food recommendation system could handle the service variance better. To be specific, the rating in the centralized food review sites often just reflects a single experience of the reviewer. However, the service of the restaurant may change over time, which reduces the information value contained in the rating. For the localized food recommendation system, the ratings keep being updated since people revisit the local restaurants frequently. Considering this effect, if a user has not been a local restaurant for a relatively long time, he would also require an updated reputation score from the localized recommendation system. 

Since the personal taste is considered as privacy by many people and the recommendation tasks could appear in different regions frequently and concurrently, to get the local reputation score, we should utilize the proposed framework to conduct the SVD-based collaborative filtering as described in \cite{sarwar2000application}. The detailed procedures are as follows:
\begin{itemize}
	\item\emph{Step-1. }The user $c$ uploads his rating vector to FD with his mobile phone and informs FD that he is interested in restaurant $p$. FD collects the rating vectors from other users inside this region. Note that to remove sparsity, each user fills in the ratings of unknown restaurants with his average rating.
	\item\emph{Step-2. }FD uploads randomized data to SD$_{d}$. SD$_{d}$, SD$_{u}$ and SD$_{v}$ conduct some extra steps for normalizing the data matrix and perform SVD to get $\mathbb{U}$, $\mathbb{V}$ and $\Sigma$. The extra normalization steps of SD$_{d}$ and SD$_{u}$ are the same as the extra steps introduced in the above "Abnormal Detection" application. Note that different with the "Abnormal Detection" application, the recommendation application requires SD$_d$ to send the $(\mathbb{A}^{'})^{T}\cdot \mathbb{A}^{'}$ to SD$_v$ as well. Since the extra normalization steps needed by SD$_v$ could be straightforward derived from the steps of SD$_{u}$, we omit the detailed description of the extra steps of SD$_v$ here.
	\item\emph{Step-3. }SD$_{u}$ and SD$_{v}$ reduce $\mathbb{U}$, $\mathbb{V}$ and $\Sigma$ to $k$ dimension, and compute $\mathbb{U}_k$$\Sigma_k^{\frac{1}{2}}$ and $\Sigma_k^{\frac{1}{2}}$$\mathbb{V}_k^{T}$ respectively, i.e. SD$_{u}$ holds $\mathbb{U}_k$$\Sigma_k^{\frac{1}{2}}$ and SD$_{v}$ holds $\Sigma_k^{\frac{1}{2}}$$\mathbb{V}_k^{T}$. Let $\mathbb{U}_k$$\Sigma_k^{\frac{1}{2}}(c)$ denote the row of $\mathbb{U}_k$$\Sigma_k^{\frac{1}{2}}$ which contains the information of user $c$ and $\Sigma_k^{\frac{1}{2}}$$\mathbb{V}_k^{T}(p)$ denote the column of $\Sigma_k^{\frac{1}{2}}$$\mathbb{V}_k^{T}$ which contains the information of restaurant $p$. The reputation score of restaurant $p$ for user $c$ is computed as $\mathbb{U}_k$$\Sigma_k^{\frac{1}{2}}(c)\Sigma_k^{\frac{1}{2}}$$\mathbb{V}_k^{T}(p)$. Note that we use z-scores for normalization, so we do not need to add the user average back as in \cite{sarwar2000application}.
	\item\emph{Step-4. }SD$_{u}$ and SD$_{v}$ send $\mathbb{U}_k$$\Sigma_k^{\frac{1}{2}}(c)$ and $\Sigma_k^{\frac{1}{2}}$$\mathbb{V}_k^{T}(p)$ to SD$_d$ for reputation score computation. To prevent SD$_d$ from inferring information, SD$_{u}$ and SD$_{v}$ randomize $\mathbb{U}_k$$\Sigma_k^{\frac{1}{2}}(c)$ and $\Sigma_k^{\frac{1}{2}}$$\mathbb{V}_k^{T}(p)$ with $W$ and $S$ respectively. For example, let $u_i$ denote the $i$th entry of $\mathbb{U}_k$$\Sigma_k^{\frac{1}{2}}(c)$, SD$_{u}$ randomizes it as $u_i+z_iW+r_iS$.
	\item\emph{Step-5. }SD$_d$ multiplies the two randomized vectors and sends the result ${score_p}^{'}$ to FD. Since FD knows $W$ and $S$, it could recover the reputation score as $score_p = {score_p}^{'} \bmod S \bmod W$ and send $score_p$ to user $c$.  
\end{itemize}
From the above description, it could be shown that the proposed framework could be utilized to compose a recommendation system without any difficulties. Also, note that in the above example, the server does not participate in the process, which means the proposed framework completes all the workload in the edge tier.

\subsection{Data Compression}
Since SVD could provide the best lower rank approximations of
original data matrix, it is largely applied for compressing data. The data compression of SVD is quite straightforward:  the matrix $\Sigma$ chooses $k$ largest diagonal values and sets the rest as 0. $\mathbb{U}$ and $\mathbb{V}$ retain the corresponding $k$ columns accordingly, then the matrix $\mathbb{U}_k \Sigma_k \mathbb{V}_k^{T}$ is the best rank $k$ approximations of original matrix. In the proposed framework, SD$_u$ and SD$_v$ could simply upload the trunked 
$\mathbb{U}_k$, $\Sigma_k$ and $\mathbb{V}_k$ to server to achieve this purpose.

Besides the lower rank approximation, the eigenvectors from $\mathbb{U}$ could be utilized to perform dimension reduction which is also a kind of data compression. In the framework, SD$_u$ could randomize the eigenvectors corresponding to the first several largest singular values. Then SD$_d$ could perform scalar product between the randomized user data and the received randomized eigenvectors. After the results of scalar product are recovered by SD$_u$, it could upload the dimension-reduced  data to server.
\section{Related Works}
\label{sec8}
In literature, there are a few works which are related to privacy-preserving SVD computation. Polat et al. \cite{polat2005svd} proposed a SVD-based collaborative filtering scheme in which the data privacy is protected by randomized perturbation. However, their scheme has been proven unsecure by \cite{zhang2006deriving}. Note that the randomization in this work does not have the feature of the randomized perturbation in \cite{polat2005svd}. Thus, the technique in \cite{zhang2006deriving} is infeasible for our work. Canny et al. \cite{canny2002collaborative} proposed a collaborative filtering scheme which achieves the SVD computation with privacy-preserving. However, their scheme is specifically designed for the recommendation application.
Han et al. \cite{han2009privacy} proposed a secure protocol for SVD computation. However, their scheme could only support the computation between two parties. Hegedűs et al. \cite{hegedHus2014fully} proposed a private SVD computation for low rank approximation in distributed P2P systems. Compared to our work, the works in \cite{canny2002collaborative, han2009privacy, hegedHus2014fully} have limited applications and require considerable iterations for convergence which brings heavy overhead. Duan et al. \cite{Duan:2010:PPL:1929820.1929839} proposed a privacy-preserving framework which supports the computation of the learning algorithms which could be expressed as iterative form. Their work could support many learning algorithms while also requiring multiple rounds for the convergence of algorithms, which brings considerable overhead. For example, their scheme needs 83 minutes to compute the SVD for the Enron Email Data set which is a $150\times 150$ matrix while our work would need $N_C =$ 15 ciphertexts to aggregate the 150-dimensional data for each of the 150 users and only takes 499 seconds in total. Note that the evaluation in \cite{Duan:2010:PPL:1929820.1929839} sums up the computation time for all users even the computation of each user is actually performed concurrently. For fair comparison, our evaluation also accumulates the computation time of all users. Thus, the overall time cost is $4\times 55.493\times 150\times 15 = 499437$ milliseconds.

\section{Conclusions}
\label{sec9}
In this paper, a flexible fog computing framework for privacy-preserving SVD computation has been proposed. The framework divides the SVD calculation into two eigenvector decomposition operations and distributes the two tasks to different fog devices. The security analysis shows that the user data privacy is preserved during transmission, aggregation and eigenvector decomposition. The possible attacks from the second layer fog devices are also analyzed and the resistance of the framework against the potential attacks is discussed. The performance analysis has indicated the capacity of the framework and shows that the data dimension is the most important factor influencing the efficiency of the system. Moreover, three IoT applications are given as examples to demonstrate that the proposed framework is flexible enough to adapt to different applications. Compared with the existing works, our framework could support large scope of applications with relatively small resource consumption.

\bibliographystyle{IEEEtran}
\bibliography{IoT}

% Generated by IEEEtran.bst, version: 1.14 (2015/08/26)
\begin{thebibliography}{10}
\providecommand{\url}[1]{#1}
\csname url@samestyle\endcsname
\providecommand{\newblock}{\relax}
\providecommand{\bibinfo}[2]{#2}
\providecommand{\BIBentrySTDinterwordspacing}{\spaceskip=0pt\relax}
\providecommand{\BIBentryALTinterwordstretchfactor}{4}
\providecommand{\BIBentryALTinterwordspacing}{\spaceskip=\fontdimen2\font plus
\BIBentryALTinterwordstretchfactor\fontdimen3\font minus
  \fontdimen4\font\relax}
\providecommand{\BIBforeignlanguage}[2]{{%
\expandafter\ifx\csname l@#1\endcsname\relax
\typeout{** WARNING: IEEEtran.bst: No hyphenation pattern has been}%
\typeout{** loaded for the language `#1'. Using the pattern for}%
\typeout{** the default language instead.}%
\else
\language=\csname l@#1\endcsname
\fi
#2}}
\providecommand{\BIBdecl}{\relax}
\BIBdecl

\bibitem{Bradley14}
\BIBentryALTinterwordspacing
J.~Bradley, J.~Barbier, and D.~Handler, ``Embracing the internet of everything
  to capture your share of \$14.4 trillion,'' 2014. [Online]. Available:
  \url{http://www.cisco.com/c/dam/en_us/about/ac79/docs/innov/IoE_Economy.pdf}
\BIBentrySTDinterwordspacing

\bibitem{McKinsey2015}
\BIBentryALTinterwordspacing
J.~Manyika, M.~Chui, P.~Bisson, J.~Woetzel, R.~Dobbs, J.~Bughin, and D.~Aharon.
  (2015, June) Unlocking the potential of the internet of things. McKinsey
  Global Institute. [Online]. Available:
  \url{http://www.mckinsey.com/business-functions/business-technology/our-insights/the-internet-of-things-the-value-of-digitizing-the-physical-world}
\BIBentrySTDinterwordspacing

\bibitem{Cisco2014}
\BIBentryALTinterwordspacing
(2014, Jan) Cisco delivers vision of fog computing to accelerate value from
  billions of connected devices. Cisco. Press release. [Online]. Available:
  \url{http://newsroom.cisco.com/release/1334100/Cisco-
  Delivers-Vision-of-Fog-Computing-to-Accelerate-Value-from-Billionsof-Connected-Devices-utm-medium-rss}
\BIBentrySTDinterwordspacing

\bibitem{ide2004eigenspace}
T.~Id{\'e} and H.~Kashima, ``Eigenspace-based anomaly detection in computer
  systems,'' in \emph{Proceedings of the tenth ACM SIGKDD international
  conference on Knowledge discovery and data mining}.\hskip 1em plus 0.5em
  minus 0.4em\relax ACM, 2004, pp. 440--449.

\bibitem{lee2013anomaly}
Y.-J. Lee, Y.-R. Yeh, and Y.-C.~F. Wang, ``Anomaly detection via online
  oversampling principal component analysis,'' \emph{IEEE Transactions on
  Knowledge and Data Engineering}, vol.~25, no.~7, pp. 1460--1470, 2013.

\bibitem{sarwar2000application}
B.~Sarwar, G.~Karypis, J.~Konstan, and J.~Riedl, ``Application of
  dimensionality reduction in recommender system-a case study,'' DTIC Document,
  Tech. Rep., 2000.

\bibitem{billsus1998learning}
D.~Billsus and M.~J. Pazzani, ``Learning collaborative information filters.''
  in \emph{Icml}, vol.~98, 1998, pp. 46--54.

\bibitem{kalman1996singularly}
D.~Kalman, ``A singularly valuable decomposition: the svd of a matrix,''
  \emph{The college mathematics journal}, vol.~27, no.~1, pp. 2--23, 1996.

\bibitem{bonomi2012fog}
F.~Bonomi, R.~Milito, J.~Zhu, and S.~Addepalli, ``Fog computing and its role in
  the internet of things,'' in \emph{Proceedings of the first edition of the
  MCC workshop on Mobile cloud computing}.\hskip 1em plus 0.5em minus
  0.4em\relax ACM, 2012, pp. 13--16.

\bibitem{paillier1999public}
P.~Paillier, ``Public-key cryptosystems based on composite degree residuosity
  classes,'' in \emph{International Conference on the Theory and Applications
  of Cryptographic Techniques}.\hskip 1em plus 0.5em minus 0.4em\relax
  Springer, 1999, pp. 223--238.

\bibitem{golub2012matrix}
G.~H. Golub and C.~F. Van~Loan, \emph{Matrix computations}.\hskip 1em plus
  0.5em minus 0.4em\relax JHU Press, 2012, vol.~3.

\bibitem{sang2009privacy}
Y.~Sang, H.~Shen, and H.~Tian, ``Privacy-preserving tuple matching in
  distributed databases,'' \emph{Knowledge and Data Engineering, IEEE
  Transactions on}, vol.~21, no.~12, pp. 1767--1782, 2009.

\bibitem{zhong2007privacy}
S.~Zhong, ``Privacy-preserving algorithms for distributed mining of frequent
  itemsets,'' \emph{Information Sciences}, vol. 177, no.~2, pp. 490--503, 2007.

\bibitem{lu2012eppa}
R.~Lu, X.~Liang, X.~Li, X.~Lin, and X.~S. Shen, ``Eppa: An efficient and
  privacy-preserving aggregation scheme for secure smart grid communications,''
  \emph{Parallel and Distributed Systems, IEEE Transactions on}, vol.~23,
  no.~9, pp. 1621--1631, 2012.

\bibitem{Goethals2004}
B.~Goethals, S.~Laur, H.~Lipmaa, and T.~Mielikäinen, ``On private scalar
  product computation for privacy-preserving data mining,'' in
  \emph{International Conference on Information Security and Cryptology}.\hskip
  1em plus 0.5em minus 0.4em\relax Springer Berlin Heidelberg, 2004, pp.
  104--120.

\bibitem{Lindell2000}
L.~Y. and P.~B., ``Privacy preserving data mining,'' in \emph{Annual
  International Cryptology Conference}.\hskip 1em plus 0.5em minus 0.4em\relax
  Springer Berlin Heidelberg, 2000, pp. 36--54.

\bibitem{kamvar2003eigentrust}
S.~D. Kamvar, M.~T. Schlosser, and H.~Garcia-Molina, ``The eigentrust algorithm
  for reputation management in p2p networks,'' in \emph{Proceedings of the 12th
  international conference on World Wide Web}.\hskip 1em plus 0.5em minus
  0.4em\relax ACM, 2003, pp. 640--651.

\bibitem{boneh2004short}
D.~Boneh, B.~Lynn, and H.~Shacham, ``Short signatures from the weil pairing,''
  \emph{Journal of cryptology}, vol.~17, no.~4, pp. 297--319, 2004.

\bibitem{nymann1972probability}
J.~Nymann, ``On the probability that k positive integers are relatively
  prime,'' \emph{Journal of Number Theory}, vol.~4, no.~5, pp. 469--473, 1972.

\bibitem{wagner2002generalized}
D.~Wagner, ``A generalized birthday problem,'' in \emph{Annual International
  Cryptology Conference}.\hskip 1em plus 0.5em minus 0.4em\relax Springer,
  2002, pp. 288--304.

\bibitem{polat2005svd}
H.~Polat and W.~Du, ``Svd-based collaborative filtering with privacy,'' in
  \emph{Proceedings of the 2005 ACM symposium on Applied computing}.\hskip 1em
  plus 0.5em minus 0.4em\relax ACM, 2005, pp. 791--795.

\bibitem{zhang2006deriving}
S.~Zhang, J.~Ford, and F.~Makedon, ``Deriving private information from randomly
  perturbed ratings,'' in \emph{Proceedings of the 2006 SIAM International
  Conference on Data Mining}.\hskip 1em plus 0.5em minus 0.4em\relax SIAM,
  2006, pp. 59--69.

\bibitem{canny2002collaborative}
J.~Canny, ``Collaborative filtering with privacy,'' in \emph{Security and
  Privacy, 2002. Proceedings. 2002 IEEE Symposium on}.\hskip 1em plus 0.5em
  minus 0.4em\relax IEEE, 2002, pp. 45--57.

\bibitem{han2009privacy}
S.~Han, W.~K. Ng, and S.~Y. Philip, ``Privacy-preserving singular value
  decomposition,'' in \emph{2009 IEEE 25th International Conference on Data
  Engineering}.\hskip 1em plus 0.5em minus 0.4em\relax IEEE, 2009, pp.
  1267--1270.

\bibitem{hegedHus2014fully}
I.~Heged{\H{u}}s, M.~Jelasity, L.~Kocsis, and A.~A. Bencz{\'u}r, ``Fully
  distributed robust singular value decomposition,'' in \emph{Peer-to-Peer
  Computing (P2P), 14-th IEEE International Conference on}.\hskip 1em plus
  0.5em minus 0.4em\relax IEEE, 2014, pp. 1--9.

\bibitem{Duan:2010:PPL:1929820.1929839}
Y.~Duan, J.~Canny, and J.~Zhan, ``P4p: Practical large-scale privacy-preserving
  distributed computation robust against malicious users,'' in
  \emph{Proceedings of the 19th USENIX Conference on Security}, ser. USENIX
  Security'10.\hskip 1em plus 0.5em minus 0.4em\relax Berkeley, CA, USA: USENIX
  Association, 2010, pp. 14--14.

\end{thebibliography}
%
% <OR> manually copy in the resultant .bbl file
% set second argument of \begin to the number of references
% (used to reserve space for the reference number labels box)
%\begin{thebibliography}{1}

%\bibitem{IEEEhowto:kopka}
%H.~Kopka and P.~W. Daly, \emph{A Guide to \LaTeX}, 3rd~ed.\hskip 1em plus
 % 0.5em minus 0.4em\relax Harlow, England: Addison-Wesley, 1999.

%\end{thebibliography}

% biography section
% 
% If you have an EPS/PDF photo (graphicx package needed) extra braces are
% needed around the contents of the optional argument to biography to prevent
% the LaTeX parser from getting confused when it sees the complicated
% \includegraphics command within an optional argument. (You could create
% your own custom macro containing the \includegraphics command to make things
% simpler here.)
%\begin{IEEEbiography}[{\includegraphics[width=1in,height=1.25in,clip,keepaspectratio]{mshell}}]{Michael Shell}
% or if you just want to reserve a space for a photo:

% You can push biographies down or up by placing
% a \vfill before or after them. The appropriate
% use of \vfill depends on what kind of text is
% on the last page and whether or not the columns
% are being equalized.

%\vfill

% Can be used to pull up biographies so that the bottom of the last one
% is flush with the other column.
%\enlargethispage{-5in}

% that's all folks
\end{document}